# High-speed and high-gain graphene photovoltaic phototransistor gated by a van der Waals heterojunction

Yihan Yin[†,#], Jiayi Zhang[†,#], Xiaolong Zhang[†], Jiongtao Zhang[†], Liang Liu[†], Haiya Ma[†], Xiaoguang Luo[†,*], Xuetao Gan[‡,*]

[†] School of Integrated Circuits and Microelectronics, State Key Laboratory of Flexible Electronics & School of Flexible Electronics, Northwestern Polytechnical University, Xi'an 710129, China

[‡] School of Physical Science and Technology, Northwestern Polytechnical University, Xi'an 710129, China

**ABSTRACT:** Two-dimensional (2D) material-based phototransistors offer a unique combination of optical sensing, signal amplification, and logic operation within a single device, yet fundamentally suffering from an inherent gain-speed trade-off. Here, we demonstrate a 2D photovoltaic phototransistor that overcomes this limitation using a $MoS_2/PtSe_2$ heterojunction to gate a graphene channel. The ultrafast photovoltaic effect in the heterojunction enables charge separation, yielding ultrahigh photoconductive gain (up to $10^8$) in graphene channel via interfacial gating. Besides, the response time (below the instrumental resolution of 550 ns) is governed by carrier transit in graphene channel, enabling simultaneous high speed and high gain. Moreover, broadband photodetection from visible to near-infrared is enabled by the optical properties of the $MoS_2/PtSe_2$ heterojunction, with the detectivity exceeding $10^{11}$ Jones. These results establish a new paradigm for high-performance 2D phototransistors by harnessing photovoltaic and photogating effects to overcome the classical gain-speed trade-off.

Since the landmark discovery of graphene (Gr) in 2004,[1] two-dimensional (2D) materials and their van der Waals heterostructures have emerged as leading candidates for next-generation optoelectronic devices. Their atomic thickness, high carrier mobility, tunable bandgaps, mechanical flexibility, and strong light-matter interaction offer distinct advantages over conventional semiconductors.[2-6] 2D material-based phototransistors can simultaneously enable optical sensing, signal amplification, and logic switching within a single device, thereby significantly simplifying and miniaturizing the layout of CMOS sensing systems. Upon illumination, photogenerated electron-hole pairs in the channel are separated and transported under an applied bias voltage. In the absence of trapping effects, the response speed is ultimately limited by the carrier transit time ($\tau_{\mathrm{tr}}$), and no photoconductive gain ($G$) is produced. In this regime, the responsivity is fundamentally bounded by $R_0 = \lambda/1240$ A/W ($\lambda$ is the light wavelength in nanometers). For instance, graphene phototransistors can sustain high-speed photoresponse up to 40 GHz due to their ultrahigh carrier mobility, yet exhibit a low responsivity of only 0.5 mA/W.[7, 8] Once one type of photogenerated carrier (electron or hole) is trapped within or near the channel, the opposite type carrier in the channel can recirculate multiple times during the trap lifetime ($\tau_{\mathrm{life}}$). This photogating (PG) effect yields a photoconductive gain $G = \tau_{\mathrm{life}}/\tau_{\mathrm{tr}}$, enabling high responsivity by a prolonged carrier lifetime. Although high responsivity is a key advantage in phototransistors, it typically comes at the expense of response speed, as illustrated in Figures 1a and 1b. A representative example is the quantum dot/Gr hybrid phototransistor, which achieves an ultrahigh responsivity of $10^8$ A/W but suffers from a slow response time of around 100 s.[9]

To overcome this fundamental trade-off in phototransistors, the introduction of built-in electric field has emerged as a promising way to improve the response speed without sacrificing gain.[10] In 2017, Adinolfi and Sargent[11] proposed the photovoltaic phototransistor (PVPT), a device architecture that integrates both photovoltaic (PV) and PG effects. By leveraging the PV effect, PVPT exhibits the capacity to achieve higher gain at faster speed. Since then, numerous approaches based on 2D materials have been explored to realize this objective. Representative designs incorporating in-plane[12, 13] or out-plane[14-16] built-in electric field have successfully reduced the response time to the millisecond or even microsecond range while maintaining high gain. However, achieving sub-microsecond response time remains challenging, likely due to the prolonged charge interactions within the device channel. A concept of interfacial gating, as introduced by Ni et al.,[17] offers a compelling alternative to further push the speed limit. By separating the channel and the photovoltaic structure with an insulator layer, the prolonged interaction between photogenerated and transported carriers is effectively suppressed. Given that the intrinsic response time of an isolated PV 2D heterostructure is on the order of picosecond,[18, 19] the overall response time of such PVPT device is expected to be governed primarily by the carrier transit time (typically nanosecond timescale) in the channel. Several recent studies adopting this design[17, 20-22] have demonstrated response times approaching the microsecond regime, albeit with moderate responsivities below $10^3$ A/W.

In this work, we demonstrate a high-speed and high-gain Gr PVPT based on an all-2D material architecture, gated by a $MoS_2/PtSe_2$ PV heterojunction. Combining the ultrafast PV response of $MoS_2/PtSe_2$ heterojunction with the ultrashort carrier transit time in Gr, the device achieves a high gain of $10^8$ (corresponding to a responsivity of $10^5$ A/W) and a fast response time < 550 ns. This result marks the realization of sub-microsecond response speeds while maintaining a high gain. Notably, the measured < 550 ns is conservatively estimated due to instrumentation limitations. Furthermore, the spectral response of our PVPT extends from visible to near-infrared region ascribed to the optical properties of $MoS_2/PtSe_2$ heterojunction,[23-25] with responsivity and detectivity remaining at high levels of $10^4$ A/W and $10^{11}$ Jones, respectively.

Figure 1c presents the device architecture of our PVPT, which consists of a $MoS_2/PtSe_2$/hBN van der Waals heterostructure stacked onto the channel of a Gr transistor on a $SiO_2$/Si substrate, where hBN denotes the hexagonal boron nitride. The Gr transistor ensures ultrafast transit of transported carriers, while the $MoS_2/PtSe_2$ heterojunction promises an ultrafast PV response and efficient separation of photogenerated electron-hole pairs. All 2D materials were mechanically exfoliated and deterministically transferred to the target location using PDMS stamps. Source and drain electrodes (Au/Cr) were fabricated via standard photolithography and electron-beam evaporation. Figure 1d shows a typical fabricated PVPT device. Atomic force microscopy (AFM) measurements reveal thicknesses of 1.43 nm (~2 layers), 1.73 nm (~3 layers), 10.3 nm (~30 layers), and 0.43 nm (~1 layer) for $MoS_2$, $PtSe_2$, hBN, and Gr nanosheets (Figure 1e), respectively, confirming the atomically thin nature of the constituent layers. Raman spectroscopy was performed under 532 nm excitation to assess material quality. As shown in Figure 1f, the sharp characteristic peaks verify the high crystalline quality of the heterostructure and indicate minimal impurity incorporation during fabrication. Specifically, the peaks at 1568 and 2660 $cm^{-1}$ correspond to the G and 2D modes of Gr,[26] while the peak at 1365 $cm^{-1}$ is assigned to the $E_{2g}$ mode of hBN.[27] The peaks at 178 and 207 $cm^{-1}$ are attributed to the $E_{2g}$ and $A_{1g}$ modes of 2H-phase $PtSe_2$,[28] and those at 385 and 407 $cm^{-1}$ correspond to the $E^1_{2g}$ and $A_{1g}$ modes of $MoS_2$,[29] respectively.

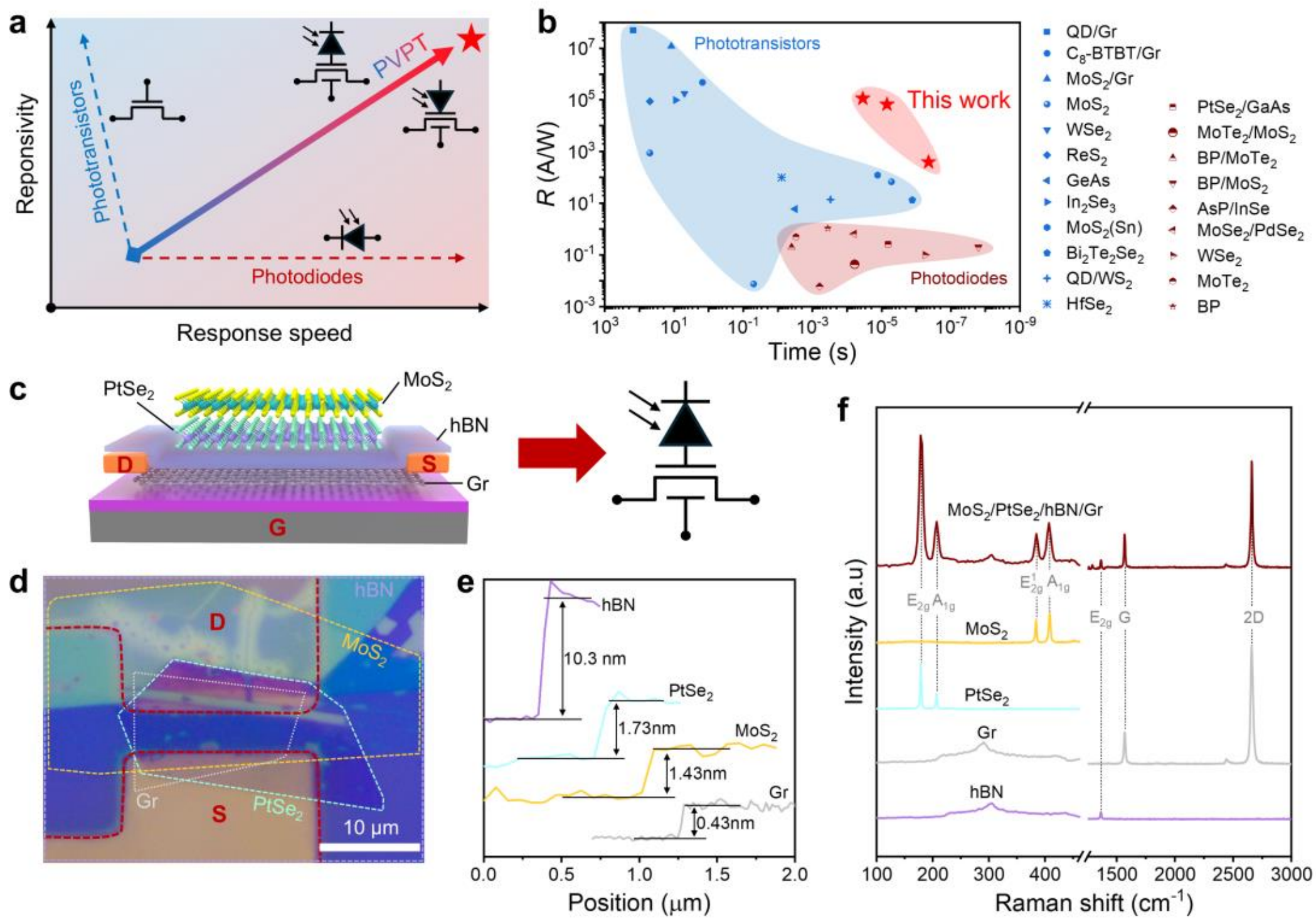

**Figure 1. Device design of PVPT with high speed and high gain.** (a) Diagram of our device benefits from both phototransistors and photodiodes. (b) Performance comparison of several typical 2D phototransistors (with uniform channel) and photodiodes (without bias) with responsivity and response time (also listed in Table S1).[9, 30-50] Response time is defined as the maximum of rise time and fall time, both extracted between 10% and 90% of the maximum photocurrent. (c) Schematical diagram of our PVPT device. (d) Optical image and (e) AFM characterizations of a fabricated device. Channel length ~4 μm and width ~15.7 μm. (f) Raman spectra of the constituent materials excited by a 532 nm laser.

The transfer characteristics (drain-source current $I_{\mathrm{DS}}$ vs. gate voltage $V_{\mathrm{G}}$) of the PVPT device were measured under 532 nm illumination with increasing laser power density $P_{\mathrm{in}}$, as shown in Figure 2a. The drain-source voltage was fixed at $V_{\mathrm{DS}} = 1$ mV throughout this study, unless otherwise specified. As $P_{\mathrm{in}}$ increases, the transfer curves shift toward negative $V_{\mathrm{G}}$, indicating photoinduced n-doping. This shift is primarily attributed to the photogating effect according to the $P_{\mathrm{in}}$-independent transconductance (Figure 2b).[51] The extracted transconductances ($g_{\mathrm{m}} = dI_{\mathrm{DS}}/dV_{\mathrm{G}}$) in the linear regime are approximately 0.2 and 0.25 μS for holes and electrons, respectively. Notably, although these values are likely underestimated due to unsaturation, they are nevertheless on the correct order of magnitude. Using the formula of $\mu = g_{\mathrm{m}} \times \frac{L}{W} \times \frac{1}{C_{\mathrm{ox}} \times V_{\mathrm{DS}}}$ (channel length/width: $L/W$, capacitance of $SiO_2$: $C_{\mathrm{ox}} = 12.11$ nF/cm$^2$), the average carrier mobilities are calculated to be $\mu_{\mathrm{h}} = 4715$ cm$^2$/V·s for holes and $\mu_{\mathrm{e}} = 5570$ cm$^2$/V·s for electrons (Figure 2c). Then the average transit times of holes and electrons are estimated to be 34.1 and 28.8 ns (Figure 2d), respectively, based on the formula of $\tau_{\mathrm{tr}} = L^2/\mu V_{\mathrm{DS}}$. These nanosecond-scale transit times allow the high-speed photoresponse of the PVPT device. The shift in transfer curves results in negative or positive photocurrent ($I_{\mathrm{ph}} = I_{\mathrm{light}} - I_{\mathrm{dark}}$, i.e., current difference between those under illumination and dark conditions) depending on the gate bias relative to the neutral voltage ($V_{\mathrm{Dirac}}$, i.e., gate voltage at Dirac point). As shown in Figure 2e, $I_{\mathrm{ph}}$ is positive for $V_{\mathrm{G}} > V_{\mathrm{Dirac}}$, and negative when $V_{\mathrm{G}} < V_{\mathrm{Dirac}}$, in excellent agreement with the photogating mechanism. With increasing $P_{\mathrm{in}}$, the neutral voltage shift ($\Delta V_{\mathrm{G}}$) increases in the inverse exponential manner and eventually saturates (Figure 2f). $\Delta V_{\mathrm{G}}$ enables the estimation of the photoinduced carrier density change via $\Delta n = C_{\mathrm{ox}} \Delta V_{\mathrm{G}}/e$, where $e$ is the elementary charge. The positive $\Delta n$ confirms an increase in electron density within the Gr channel. Furthermore, per unit time ($t = 1$ s), $\Delta n$ can be used to evaluate the quantum efficiency of PVPT device by $\eta = (\Delta n/t)/\Phi_{\mathrm{in}}$, where $\Phi_{\mathrm{in}} = P_{\mathrm{in}}/E_{\mathrm{ph}}$ (in units of m$^{-2}$s$^{-1}$) is the incident photon flux, determined by the incident power density $P_{\mathrm{in}}$ and the photon energy $E_{\mathrm{ph}}$.[52] For instance, $\Delta V_G \sim 1$ V at $P_{in} = 0.016$ mW/cm$^2$ (Figure 2f), yielding a quantum efficiency $\eta \sim 1.7‱$ (i.e., approximately 17 photoinduced carriers in Gr channel per 10000 incident photons).

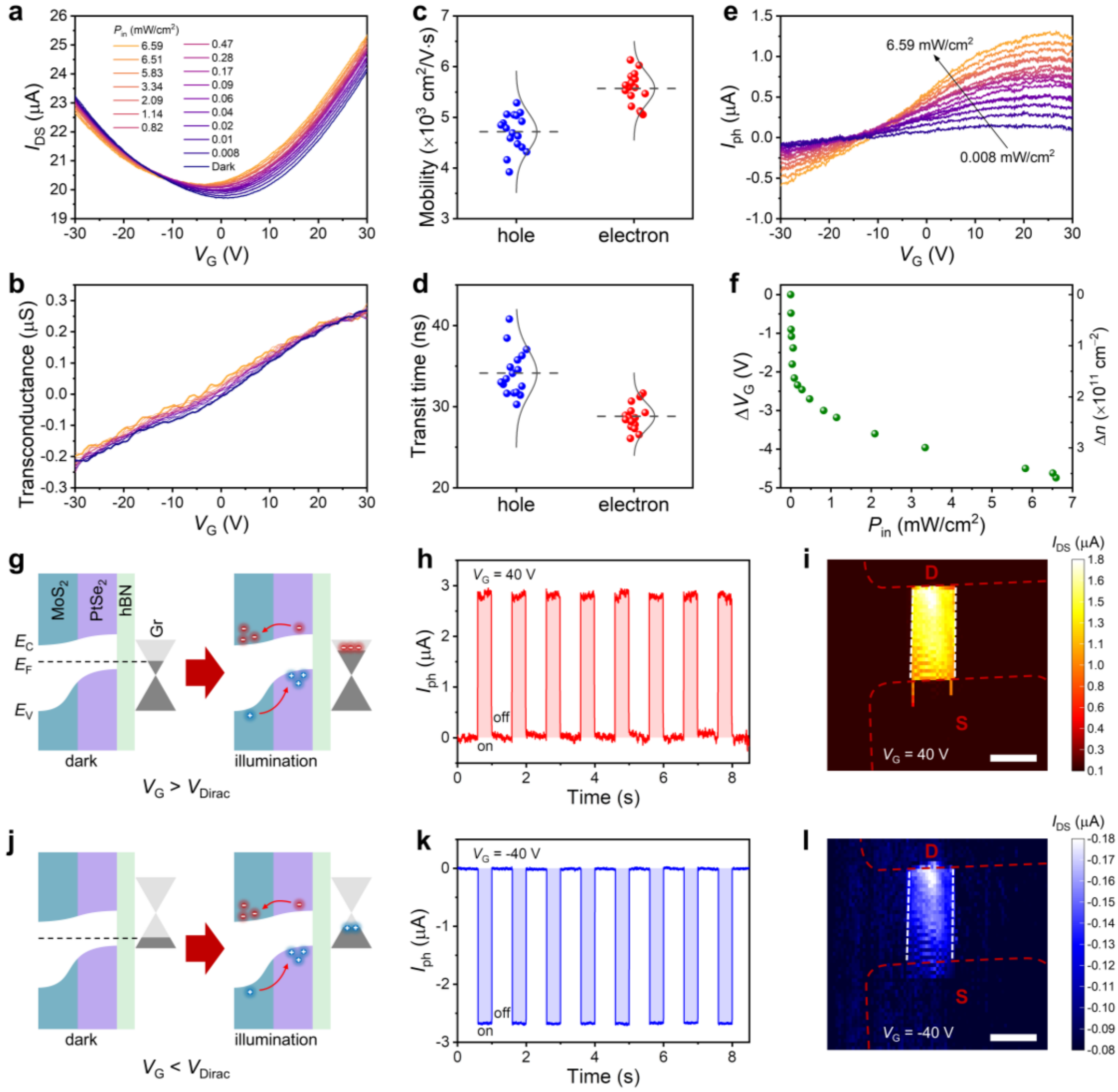

**Figure 2. Photodetection mechanism of the device.** (a) Transfer curves of the device (Figure 1d) under different 532 nm illumination at $V_{DS} = 1$ mV. (b) The corresponding transconductances. (c) Calculated mobility for holes ($V_G = -30$ V) and electrons ($V_G = 30$ V). (d) Calculated transit time. (e) Extracted photocurrent based on transfer curves. (f) Dirac point shift (left) and carrier density change (right) at different laser power density. (g) Energy band diagrams of the $MoS_2$/$PtSe_2$-gated Gr PVPT under dark and illumination conditions for electron transport ($V_G > V_{Dirac}$). (h) Positive photocurrent generated under 532 nm illumination at $V_G = 40$ V and $P_{in} = 14.5$ mW/cm$^2$. (i) Photocurrent map of a device (Figure S2) at $V_G = 40$ V. (j) Band structure diagrams of the device under dark and illumination conditions for the hole transport ($V_G < V_{Dirac}$). (k) Negative photocurrent generated under 532 nm illumination at $V_G = -40$ V and $P_{in} = 14.5$ mW/cm$^2$. (l) Photocurrent map of a device (Figure S2) at $V_G = -40$ V. The laser spot diameter is ~3 μm and power is ~0.2 μW for the photocurrent scanning measurements. Scalebar: 5 μm.

To elucidate the operating mechanism of the PVPT device, we present the energy band diagram in Figures 2g and 2j. The work function of narrow bandgap $PtSe_2$ films is positioned at −4.84 eV according to previous reports,[53] and the conduction band minimum and valence band maximum of few-layer $MoS_2$ are located at −4.3 and −5.6 eV, respectively.[54] According to the Kelvin probe force microscopy (KPFM) measurement (Figure S1), the contact potential difference of $PtSe_2$ is 260 mV larger than that of $MoS_2$, indicating the Fermi level of $PtSe_2$ is lower than $MoS_2$ (Figure S1d).[55] Consequently, a type-I or type-II band alignment is likely formed at the $MoS_2$/$PtSe_2$ heterojunction, with the built-in electric field oriented from $MoS_2$ toward $PtSe_2$. As illustrated in Figure 2g, electron-hole pairs generated in $MoS_2$ and or $PtSe_2$ under illumination are efficiently separated by this built-in electric field. Photogenerated electrons accumulate in $MoS_2$ via the PV effect, while holes accumulate in $PtSe_2$. Charge conservation is maintained in the $MoS_2$/$PtSe_2$ heterojunction if ignoring carrier tunneling through hBN layer. Once turning off the illumination, the photogenerated electrons and holes recombine on a typical picosecond timescale.[18] The accumulated holes in $PtSe_2$ induce the photogating effect in the underlying Gr channel, and further screen the electrostatic influence of electrons accumulated in the top $MoS_2$. This hole-mediated photogating increases the electron density in the Gr channel, thereby raising the Fermi level. For $V_G > V_{\mathrm{Dirac}}$, the increased electrons population lead to a positive photocurrent due to the n-type transport behavior (Figures 2g and 2h). Conversely, for $V_G < V_{\mathrm{Dirac}}$, the increase in electron population results in a negative photocurrent due to the p-type transport regime, as shown in Figures 2j and 2k. To directly visualize photocurrent generation, spatial-resolved photocurrent mapping was performed using a home-built measurement system. As shown in Figures 2i and 2l, under both $V_G > V_{\mathrm{Dirac}}$ and $V_G < V_{\mathrm{Dirac}}$ conditions, the dominating photocurrent is observed across the Gr channel region, further corroborating the photogating mechanism of the PVPT device. Since the photocurrent is predominantly generated in the overlap region between Gr channel and $MoS_2$/$PtSe_2$ heterojunction, the entire Gr channel is reasonably regarded as the effective photoactive area of the PVPT device.

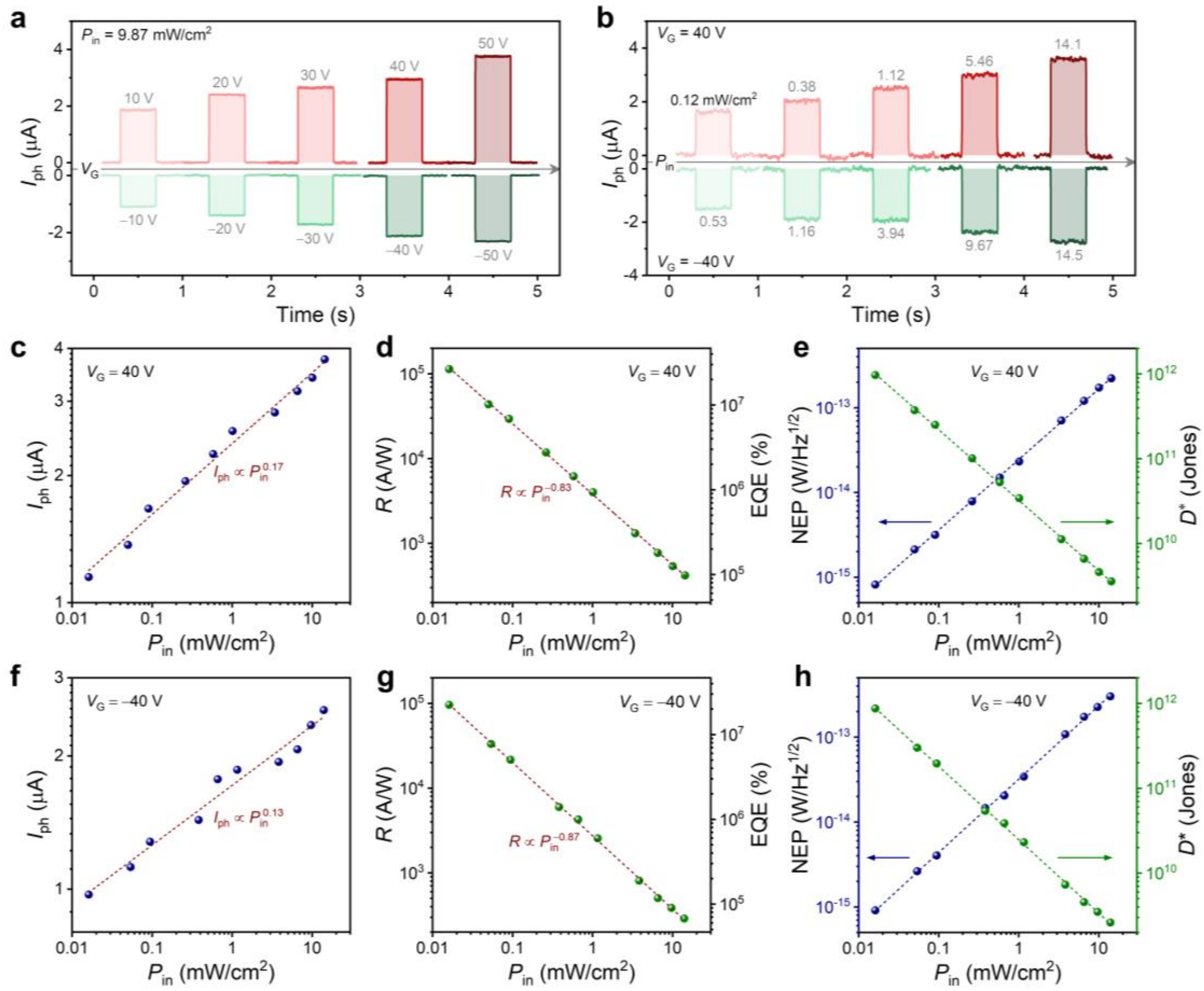

**Figure 3. Photodetection performance of the PVPT device.** (a) Transient photocurrent for different $V_{\mathrm{G}}$ at $P_{\mathrm{in}} = 9.87$ mW/cm$^2$. (b) Transient photocurrent for different $P_{\mathrm{in}}$ at $V_{\mathrm{G}} = \pm 40$ V. (c-e) $I_{\mathrm{ph}}$, $R$, $EQE$, $NEP$, $D^*$ with respect to $P_{\mathrm{in}}$ when $V_{\mathrm{G}} = 40$ V. (f-h) corresponding performances with respect to $P_{\mathrm{in}}$ when $V_{\mathrm{G}} = -40$ V.

To evaluate the photodetection performance of the PVPT device, transient photocurrent was measured under 532 nm laser excitation modulated at a frequency of 1 Hz (see experimental setup in Figure 4a). The laser spot with diameter of ~ 5 mm is used for global illumination to eliminate the photo-thermoelectric effect induced by temperature gradient arising from non-uniform illumination. Figure 3a displays the transient photocurrent at different gate voltages when $P_{\mathrm{in}} = 9.87$ mW/cm$^2$. The sharp on-off switching reveals a fast photoresponse of the device. As $V_{\mathrm{G}}$ increases from 10 to 50 V, the positive photocurrent rises monotonically to several microamperes, consistent with the trend of transconductance. A similar behavior is observed for negative photocurrent at negative gate voltages. At a fixed gate voltage (either $V_{\mathrm{G}} > V_{\mathrm{Dirac}}$ or $V_{\mathrm{G}} < V_{\mathrm{Dirac}}$), both positive and negative photocurrent increase with increasing $P_{\mathrm{in}}$, as shown in Figures 3b, 3c, and 3f. The relationship between photocurrent and laser power density follows a power law of $I_{\text{ph}} \propto P_{\text{in}}^{\gamma}$ with $\gamma < 1$, i.e., a typical sublinear dependence characteristic of photogating effect.[51] For comparison, a phototransistor with a pure Gr channel was measured under the same illumination conditions, and no observable photocurrent was detected even at a high power density of 8.7 mW/cm$^2$ (Figure S3), highlighting the crucial role of the $MoS_2$/$PtSe_2$ heterojunction in the photoresponse. The PVPT is a floating-gate electronic device, as evidenced by the memory window in the dual-sweeping transfer curves (Figure S4a). Fortunately, this memory effect does not affect the stable photoresponse in the visible and near-infrared bands, because of the high interface barrier.[56] Photocarriers generated in the $MoS_2$/$PtSe_2$ heterojunction do not spontaneously tunnel into the Gr channel, confirmed by the largely unchanged photocurrent and dark current (Figures S4b-S4d).

The responsivity, a key figure of merit quantifying the input-output efficiency of a photodetector,[57] was calculated by:

$$R = \frac{I_{\text{ph}}}{P_{\text{in}} \cdot A} \tag{1}$$

where $A$ is the effective photoactive area (i.e., the Gr channel area of our PVPT device), and $A = 62.8\ \mu\text{m}^2$ for the device of Figure 1d. The obtained responsivity decays exponentially with increasing $P_{\text{in}}$, which is well-fitted by $R \propto P_{\text{in}}^{1-\gamma}$. As $V_{\text{G}} = 40$ V, the largest measured responsivity reaches $1.14 \times 10^5$ A/W at $P_{\text{in}} = 0.016$ mW/cm$^2$. It is worth noting that even higher $R$ can be expected under weaker illumination, according to previous reports.[9, 52] To eliminate the quantization error in $P_{\text{in}}$, the focused laser spot (diameter ~ 3 μm) was also used for transient photocurrent measurements (Figure S6), and the maximum responsivity reached a comparable level of $1.3 \times 10^5$ A/W.

Such an exceptionally high responsivity verifies the presence of ultrahigh photoconductive gain. For sensitized photodetectors, photoconductive gain is commonly defined as the number of detected charge carriers per effective incident photon,[58] given by

$$G = \frac{I_{\mathrm{ph}}/e}{(\Phi_{\mathrm{in}}A)\eta} = \frac{EQE}{\eta} = R\frac{hc}{e\lambda\eta} \quad (2)$$

where $EQE = (I_{\mathrm{ph}}/e)/(\Phi_{\mathrm{in}}A) = Rhc/e\lambda$ is the external quantum efficiency, $h$ is the Planck constant, and $c$ is the speed of light, and quantum efficiency $\eta$ is defined as the ratio of photoinduced carriers in Gr channel to incident photons.[52] The maximum $EQE$ is calculated as $2.65 \times 10^7\%$, yielding a corresponding gain $G = 1.55 \times 10^8$ when taking the quantum efficiency of $\eta \sim 1.7‰$, as summarized in Table S2.

The sensitivity of a photodetector is commonly assessed using the figures of merit: noise equivalent power ($NEP$) or detectivity ($D^*$). $NEP$ (in units of $\mathrm{W/Hz^{1/2}}$) represents the minimum detectable optical power at unity signal-to-noise ratio and is defined as:[59]

$$NEP = \frac{i_{\mathrm{n}}}{R} \quad (3)$$

where the noise current $i_{\mathrm{n}}$ (in $\mathrm{A/Hz^{1/2}}$) is the random root-mean-square fluctuation in current per unit bandwidth. Generally, for imaging applications, the photodetector is typically regarded as a low-pass filter for noise, and the noise current can be calculated as $i_{\mathrm{n}} = \sqrt{\int_0^{\Delta f} S(f)/\Delta f df}$, where $S(f)$ is the spectral noise density (in $\mathrm{A^2/Hz}$), $f$ is the operating frequency, and $\Delta f$ is the noise bandwidth. When evaluating a single photodetector with an external narrowband filter centered at the operating frequency, particularly for a normalized bandwidth of 1 Hz, the noise current simplifies to $i_{\mathrm{n}} = \sqrt{S(f)}$. In fact, $i_{\mathrm{n}} = \sqrt{S(f)}$ can be used to account for the noise current when $\Delta f$ is sufficiently small, leading to the simplified expression $NEP = \sqrt{S(f)}/R$.[60] $D^*$ (in $\mathrm{cm{\cdot}Hz^{1/2}/W}$ or Jones) is the normalized signal-to-noise ratio over the active area and bandwidth through

$$D^* = \frac{\sqrt{A}}{NEP} \quad (4)$$

which can be simplified as $D^* = R\sqrt{A/S(f)}$ for small $\Delta f$. Figure S7 shows the spectral noise density of the device measured under dark conditions. $1/f$ noise dominates for the given gate voltages, consistent with the behavior of conventional field-effect transistors. At the measuring frequency of 1 Hz, $S(1\ \mathrm{Hz}) = 8.62 \times 10^{-21}$ $\mathrm{A^2/Hz}$ for $V_G = 40$ V and $S(1\,\text{Hz}) = 7.65 \times 10^{-21}$ A$^2$/Hz for $V_G = -40$ V. Then, we calculated $NEP$ and $D^*$ as functions of $P_{\text{in}}$ at $V_G = 40$ V (Figure 3e), with the minimum $NEP = 0.82$ fW/Hz$^{1/2}$ and maximum $D^* = 9.7 \times 10^{11}$ Jones. Comparable performance is obtained for the negative photoresponse regime at $V_G = -40$ V (Figures 3g and 3h), with the maximum $R$ of $0.96 \times 10^5$ A/W, minimum $NEP$ of 0.91 fW/Hz$^{1/2}$, and maximum $D^*$ of $8.7 \times 10^{11}$ Jones.

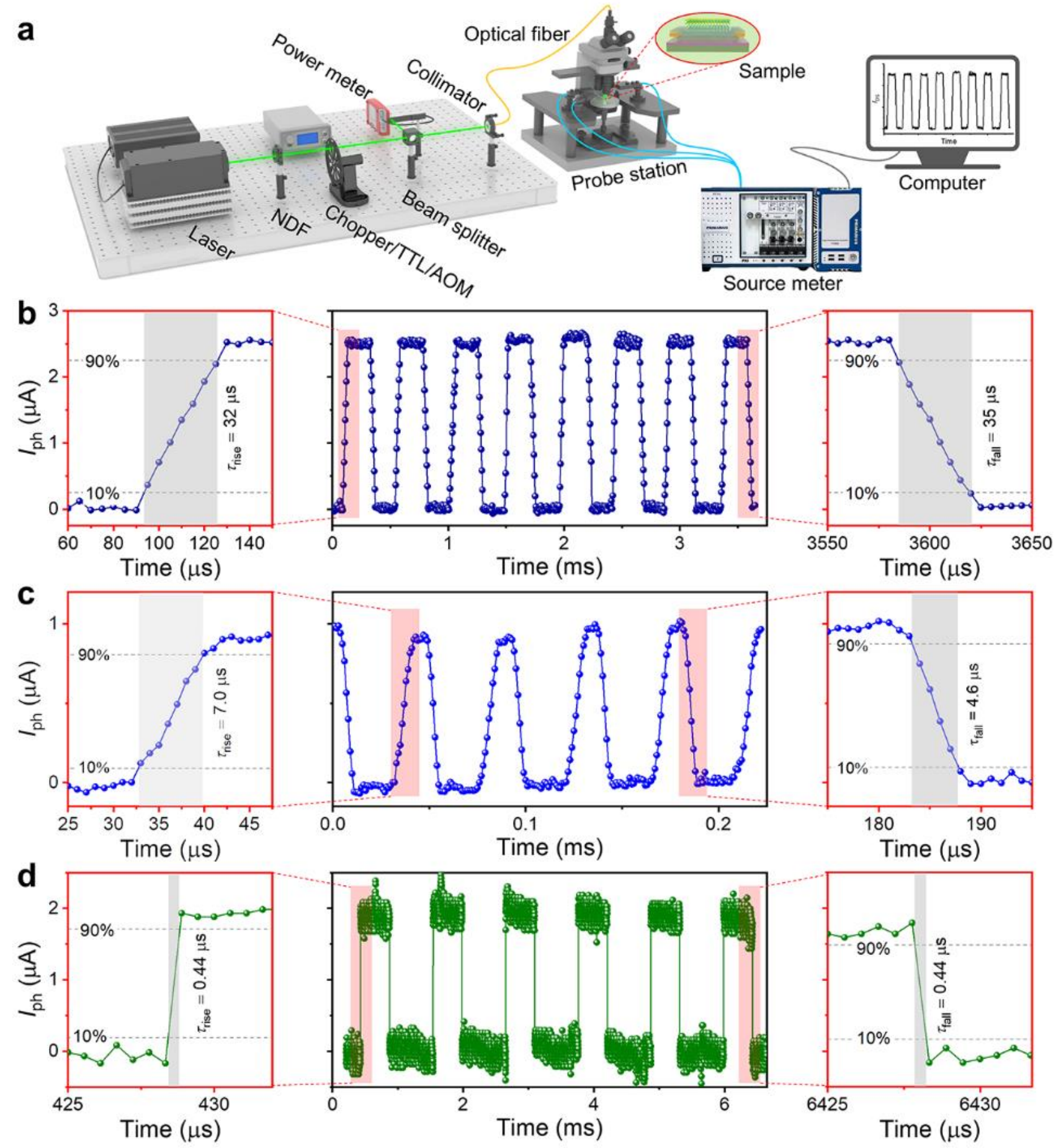

**Figure 4. Photoresponse speed of the PVPT device under 532 nm illumination.** (a) Schematic diagram of the measuring setup. Laser spot diameter: 5 mm. NDF: neutral density filter, TTL: transistor–transistor logic, AOM: acousto–optic modulator. (b) Transient photocurrent and response time as the laser ($P_{\text{in}} = 19.89$ mW/cm$^2$) is modulated by a chopper ($f = 2500$ Hz). (c) Results of a device (Figure S8a) as the laser ($P_{\text{in}} = 0.03$ mW/cm$^2$) is modulated by TTL ($f = 25$ kHz), where $V_G = 45$ V and $V_{DS} = 0.01$ V. (d) Results of a device (Figure S8b) as the laser ($P_{\text{in}} = 6.45$ mW/cm$^2$) is modulated by an AOM ($f = 1000$ Hz). $V_G = 40$ V is applied for both (b) and (d).

In addition to ultrahigh gain, our PVPT device exhibits exceptionally high speed. As modulating the laser with a chopper (Figure 4a), the rise ($\tau_{\text{rise}}$) and fall ($\tau_{\text{fall}}$) times (measured between 10% and 90% of the maximum photocurrent) for the device (Figure 1d) were extracted as 32 and 35 μs, respectively, as shown in Figure 4b. The linear rise and fall tails persist even at high modulation frequencies, likely arising from the gradual mechanical shuttering of the chopper, suggesting an intrinsic photoresponse time substantially shorter than 30 μs. To verify the high speed, two additional devices (Figure S8) were fabricated for high-speed measurements, employing transistor-transistor logic (TTL) and acousto-optic modulator (AOM), respectively, in place of the chopper. For TTL, the photocurrent waveform was well preserved at optical modulation frequencies below 25 kHz (Figures 4c and S9). At $P_{\mathrm{in}} = 0.03$ mW/cm$^2$, $\tau_{\mathrm{rise}} = 7.0$ μs and $\tau_{\mathrm{fall}} = 4.6$ μs are extracted, accompanied by a high responsivity of $6.73\times10^4$ A/W. For AOM, the fast-switching capability (sub-100 ns) enables more accurate determination of the photoresponse time. As shown in Figure 4d, both $\tau_{\mathrm{rise}}$ and $\tau_{\mathrm{fall}}$ are extracted as 440 ns (defined between 10% and 90% of the maximum photocurrent). However, due to the minimum sampling interval (0.55 μs) of our source meter, these values only indicate that the actual response time is shorter than 550 ns (an upper bound limited by instrumentation), rather than providing a precise intrinsic measurement.

In conventional phototransistors operating via charge trapping, the gain is given by $G = \tau_{\mathrm{life}}/\tau_{\mathrm{tr}}$, where $\tau_{\mathrm{life}}$ can be approximated as the $\tau_{\mathrm{fall}}$.[51] Applying this model to positive photoresponse of our PVPT device yields a gain of less than $10^3$, which is orders of magnitude lower than the experimentally determined gain of $10^8$ based on the quantum efficiency. This marked discrepancy suggests that the ultrahigh gain is unlikely dominated by conventional charge-trapping photogating effect, but instead originates from interfacial gating of the hole accumulation in $PtSe_2$ layer. To further explore this, we performed temperature-dependent photocurrent measurements. In general, increasing temperature can promote charge trapping through more effective defect ionization, while simultaneously reducing carrier mobility via enhanced scattering. The competition of these effects often leads to a typical photocurrent that first increases and then decreases with temperature.[61, 62] In our PVPT device, however, the photocurrent remains largely stable over 10-300 K range at the specified $P_{\mathrm{in}}$ (Figure S10), showing no pronounced temperature dependence. Moreover, at room temperature, the transfer characteristics are independent of the gate voltage sweep rate (Figure S11), which further suggests that the charge trapping is not the primary operating mechanism. Although we cannot completely rule out some contribution from shallow traps, interfacial states, or other persistent trap-assisted photogating, their influence (if present) does not dominate the observed photoresponse.

Reversing the stacking order of the PV heterojunction enables the role of electron accumulation in $MoS_2$ layer. Figure S12 shows the structure of Gr PVPT device gated by $PtSe_2/MoS_2$. In this configuration, accumulated electrons reduce the electron population in Gr channel via gating effect (i.e., lowering the Fermi level), leading to negative photocurrent when $V_G > V_{Dirac}$ and positive photocurrent when $V_G < V_{Dirac}$, (Figure S13). The photodetection performances (Figure S14) are comparable to those of the $MoS_2/PtSe_2$ PVPT device, demonstrating versatility and design flexibility of the PV heterojunction.

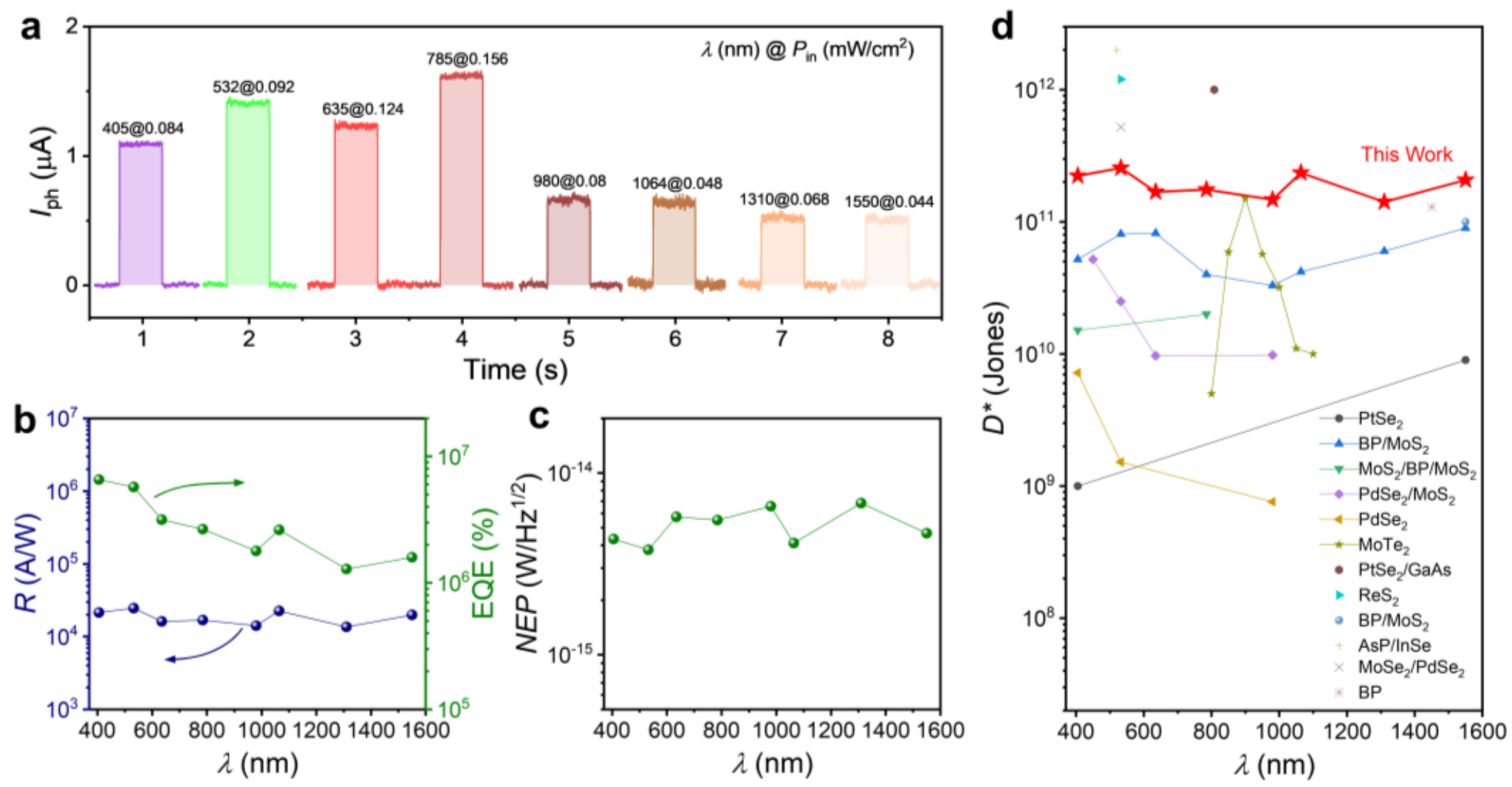

**Figure 5. Broadband photoresponse and performance benchmarking.** (a) Transient photocurrent of $MoS_2/PtSe_2$ PVPT device under illumination from 405 to 1550 nm. (b, c) Corresponding $R$, $EQE$ and $NEP$ at different wavelengths. (d) Performance comparison of detectivity with state-of-the-art 2D photodetectors.[35, 42, 43, 45, 46, 48, 49, 63, 64]

Using the transfer matrix method, we calculated the light absorption spectra (400-1600 nm) of individual flakes and complete flake stack in the PVPT device, respectively, as shown in Figure S15. For individual flakes, the response ranges are estimated to be 400-900 nm for $MoS_2$, and 400-1600 nm for both $PtSe_2$ and Gr, consistent with their respective optical bandgaps. Since light absorption in Gr does not yield appreciable photoconductive gain (Figure S3), the spectral response of the PVPT device is primarily governed by the constituent PV heterojunction. In the near-infrared region, $PtSe_2$ plays a key role in photoresponse, as further evidenced by the single-channel control experiments shown in Figure S16. Considering the absorption properties of $MoS_2$ and $PtSe_2$, our device enables broadband photodetection from at least 400 to 1600 nm, as further confirmed by the calculated results for the complete flake stack (Figure S15h). Figures 5a and S17 show the experimental photoresponse of the device from visible to near-infrared (405-1550 nm). Across all measured wavelengths, the device maintains the high responsivity of $10^4$ A/W under weak illumination, accompanied by a consistently $NEP$ at the level of 1 fW/Hz$^{1/2}$ (Figures 5b and 5c). Correspondingly, the detectivity reaches the high value of $10^{11}$ Jones, demonstrating superior performance compared to various other 2D photodetectors, as compared in Figure 5d. The absorption spectrum of $MoS_2$/$PtSe_2$ heterostructure (Figure S15h) does not directly correspond to the responsivity spectrum (Figure 17a), particularly in the near-infrared region. This slight discrepancy likely arises from nonlinear optical effects, such as second-harmonic generation in $MoS_2$ and its heterojunction[24, 25] and interlayer transition in van der Waals heterostructures,[65, 66] in addition to the absorption contribution from $PtSe_2$. To calibrate the performance at 1550 nm—an important communication wavelength—we characterized both our PVPT device and a commercial InGaAs photodiode under the same optical path with comparable $P_{\text{in}}$ levels (Figure S18). Clear on-off photocurrent switching is observed for both devices, yielding responsivities of ~ $10^4$ A/W and ~ 0.85 A/W, respectively, which confirms the high performance of our PVPT device.

In summary, we have developed all-2D photovoltaic phototransistors that simultaneously achieves ultrahigh gain and sub-microsecond response speed, to resolve a long-standing challenge in phototransistor design. By integrating a $MoS_2$/$PtSe_2$ heterojunction with a graphene transistor, the device leverages the photovoltaic effect for efficient light harvesting and rapid photoresponse, and the interfacial photogating effect for high photoconductive gain. The response time of the device is governed primarily by the carrier transit time in the graphene channel, demonstrating that higher gain can be attained without compromising speed. The measured gain of $10^8$ and response time < 550 ns represent a significant advance over conventional phototransistors, along with a high responsivity of $1.14 \times 10^5$ A/W, a low noise equivalent power of 0.82 fW/Hz$^{1/2}$, and a high detectivity of $9.7 \times 10^{11}$ Jones. These superior performance metrics are readily reproducible from device to device, and hold promise for further optimization, as summarized in Figure S19. Moreover, the device exhibits broadband photodetection from visible to near-infrared with consistently high performance. Importantly, the versatility of the photovoltaic heterojunction design offers flexibility for diverse optoelectronic applications, particularly in extending the spectral range.[67-69] With ultrahigh sensitivity, high speed, and broadband response, our photovoltaic phototransistor paves a promising way for next-generation optical communication, sensing, and imaging systems.

## ■ ASSOCIATED CONTENT

### Supporting Information

The Supporting Information is available free of charge at https://.

Detailed device fabrication and electrical characterization; performance benchmark of several typical 2D phototransistors and photodiodes; spectral noise density of the main device; optical images of devices for photocurrent scanning and high-speed testing; performance of a PVPT device gated by $PtSe_2/MoS_2$ heterostructure; broadband photoresponse of the main device (PDF)

## ■ AUTHOR INFORMATION

### Corresponding Authors

**Xiaoguang Luo** – School of Integrated Circuits and Microelectronics, State Key Laboratory of Flexible Electronics & School of Flexible Electronics, Northwestern Polytechnical University, Xi'an 710129, China; Email: iamxgluo@nwpu.edu.cn

**Xuetao Gan** – School of Physical Science and Technology, Northwestern Polytechnical University, Xi'an 710129, China; Email: xuetaogan@nwpu.edu.cn

## Authors

**Yihan Yin** – School of Integrated Circuits and Microelectronics, State Key Laboratory of Flexible Electronics & School of Flexible Electronics, Northwestern Polytechnical University, Xi'an 710129, China;

**Jiayi Zhang** – School of Integrated Circuits and Microelectronics, State Key Laboratory of Flexible Electronics & School of Flexible Electronics, Northwestern Polytechnical University, Xi'an 710129, China;

**Xiaolong Zhang** – School of Integrated Circuits and Microelectronics, State Key Laboratory of Flexible Electronics & School of Flexible Electronics, Northwestern Polytechnical University, Xi'an 710129, China;

**Jiongtao Zhang** – School of Integrated Circuits and Microelectronics, State Key Laboratory of Flexible Electronics & School of Flexible Electronics, Northwestern Polytechnical University, Xi'an 710129, China;

**Liang Liu** – School of Integrated Circuits and Microelectronics, State Key Laboratory of Flexible Electronics & School of Flexible Electronics, Northwestern Polytechnical University, Xi'an 710129, China;

**Haiya Ma** – School of Integrated Circuits and Microelectronics, State Key Laboratory of Flexible Electronics & School of Flexible Electronics, Northwestern Polytechnical University, Xi'an 710129, China;

## Author Contributions

[#] Y.Y. and Jiayi Zhang contributed equally to this work. Y.Y. and X.L. conceived the study. X.L. and X.G. supervised the research. Y.Y. and Jiayi Zhang fabricated the devices and performed the measurements. X.Z., Jiongtao Zhang, L.L., and H.M. contributed to the device fabrication. Y.Y. and X.L. cowrote the manuscript with contributions from all the authors. All authors discussed the results and contributed to writing the manuscript.

## Notes

The authors declare no competing financial interest.

## ACKNOWLEDGEMENTS

This work was supported by the National Natural Science Foundation of China (Nos. 12574459 and 12374359), the Natural Science Basic Research Program of Shaanxi (No. 2025JC-YBMS-654).

Supporting information for

# High-speed and high-gain graphene photovoltaic phototransistor gated by a van der Waals heterojunction

Yihan Yin[†,#], Jiayi Zhang[†,#], Xiaolong Zhang[†], Jiongtao Zhang[†], Liang Liu[†], Haiya Ma[†], Xiaoguang Luo[†,*], Xuetao Gan [‡,*]

[†] School of Integrated Circuits and Microelectronics, State Key Laboratory of Flexible Electronics & School of Flexible Electronics, Northwestern Polytechnical University, Xi'an 710129, China

[‡] School of Physical Science and Technology, Northwestern Polytechnical University, Xi'an 710129, China

[#] Yihan Yin and Jiayi Zhang contributed equally to this work

[*] Correspondence: Xiaoguang Luo (iamxgluo@nwpu.edu.cn) | Xuetao Gan (xuetaogan@nwpu.edu.cn)

## Experimental Methods

***Fabrication:*** All 2D materials ($MoS_2$, $PtSe_2$, hBN, and graphene) were mechanically exfoliated from commercial bulk crystals (HQ Graphene) using the Scotch-tape (3M) method. The heterostructure of photovoltaic phototransistor was fabricated via a dry-transfer technique (E1–T, MetaTest) with polydimethylsiloxane (PDMS) stamp. Graphene was first transferred onto a pre-cleaned $SiO_2$/Si substrate (heavily n-doped Si with resistivity < 0.01 Ω cm, $SiO_2$ thickness ~285 nm). The source and drain electrodes (45-nm Au/5-nm Cr) were fabricated by a maskless ultraviolet lithography system (405 nm, ATS–07–UV Litho–ACA, TuoTuo Technology) and an electron-beam evaporation system (AMOD, Angstrom Engineering Inc). A hBN nanosheet was then transferred on the graphene channel as an insulator layer. The hBN thickness greater than 5 nm is preferred to suppress Fowler-Nordheim tunneling and minimize direct tunneling of photogenerated charge carriers in the $MoS_2$/$PtSe_2$ heterojunction. Subsequently, $PtSe_2$ and $MoS_2$ nanosheets were sequentially transferred onto the hBN-covered graphene channel as the photovoltaic heterojunction. Finally, the fabricated device was annealed at 200 ℃ for 2 hours in a vacuum tube furnace (BTF–1200C–S, BEQ) to remove residual resist and reduce contact resistance.

***Characterization:*** Raman spectra of $MoS_2$, $PtSe_2$, hBN, graphene, and their heterostructure were collected by using a confocal micro-Raman system (Alpha300R, WITec) with 532 nm laser excitation (spot size ~400 nm, laser power ~0.5 mW). AFM measurements were performed using an atomic force microscope (Dimension Icon, Bruker). All electrical and optoelectrical measurements were performed at room temperature in a probe station using a semiconductor parameter analyzer (FS380 Pro, Platform Design Automation), which was also used for spectral noise density measurements with its noise mode. Photodetection performances were characterized using laser sources at wavelengths of 405, 532 (spot diameter ~5 mm), 635, 785, 980, 1064, 1310, and 1550 nm, respectively. Spatial-resolved photocurrent mapping was carried out under ambient conditions using a home–built scanning system with a focused 532 nm laser (spot diameter ~3 μm).

**Table S1.** Performance characteristics of several typical 2D phototransistors (with uniform channel) and photodiodes (without bias)[a)]

| Materials | Wavelength [nm] | $V_{DS}$ [V] | $V_G$ [V] | $R$ [A/W] | Gain | $D^*$ [Jones] | Time [s][b)] | Ref. |
|---|---|---|---|---|---|---|---|---|
| $MoS_2/PtSe_2/hBN/Gr$ | 532 | 0.001 | 40 | $1.14\times10^5$ | $1.55\times10^8$ | $9.7\times10^{11}$ | $3.5\times10^{-5}$ | This Work |
| | | | | 395 | - | - | $4.4\times10^{-7}$ | |
| | | 0.01 | 45 | $6.73\times10^4$ | - | $5.6\times10^{11}$ | $7.0\times10^{-6}$ | |
| QD/Gr | 600 | 5 | -20 | $5\times10^7$ | $1\times10^8$ | $7\times10^{13}$ | ~150 | 1 |
| $C_8$-BTBT/Gr | 355 | 0.1 | - | $4.76\times10^5$ | $1.84\times10^9$ | - | 1.5 | 2 |
| $MoS_2$/Gr | 650 | 0.1 | -10 | $1.2\times10^7$ | $10^8$ | - | 12 | 3 |
| $MoS_2$ | 561 | 8 | -70 | 880 | $2\times10^4$ | - | ~50 | 4 |
| $MoS_2$ | 550 | 1 | 50 | $7.5\times10^{-3}$ | - | - | 0.05 | 5 |
| $WSe_2$ | 650 | 2 | -60 | $1.8\times10^5$ | $3.5\times10^5$ | $3\times10^{14}$ | 5 | 6 |
| $ReS_2$ | 532 | 4 | -50 | $8.86\times10^4$ | - | $1.2\times10^{12}$ | ~50 | 7 |
| GeAs | 1600 | 2 | 60 | 6 | - | - | $3.2\times10^{-3}$ | 8 |
| $In_2Se_3$ | 640 | 0.05 | 30 | $9.8\times10^4$ | - | $3.3\times10^{13}$ | 9 | 9 |
| $MoS_2$(Sn) | 850 | -2 | - | ~123 | ~1000 | - | $1.3\times10^{-5}$ | 10 |
| $MoS_2$ | | | | ~68 | ~100 | | $5.2\times10^{-6}$ | |
| $Bi_2Te_2Se_2$ | 1550 | 1 | 0 | 13.5 | - | $6\times10^9$ | $1.3\times10^{-6}$ | 11 |
| QD/$WS_2$ | 808 | 1.5 | 2 | 14 | 21.48 | $3.9\times10^8$ | ~$3\times10^{-4}$ | 12 |
| $HfSe_2$ | 473 | 2 | 40 | ~100 | ~200 | $2.5\times10^{11}$ | $7.8\times10^{-3}$ | 13 |
| $PtSe_2$/GaAs | 808 | 0 | - | 0.26 | - | $2.5\times10^{12}$ | $6.5\times10^{-6}$ | 14 |
| $MoTe_2/MoS_2$ | 637 | 0 | - | $4.36\times10^{-2}$ | - | $1.1\times10^8$ | $6\times10^{-5}$ | 15 |
| BP/$MoTe_2$ | 532 | 0 | 30 | 0.2 | - | - | 0.004 | 16 |
| BP/$MoS_2$ | 1550 | 0 | - | 0.2 | - | $10^{11}$ | $1.6\times10^{-8}$ | 17 |
| AsP/InSe | 520 | 0 | - | 0.006 | - | - | $6.2\times10^{-4}$ | 18 |
| $MoSe_2/PdSe_2$ | 532 | 0 | - | 0.651 | - | $5.3\times10^{11}$ | $6.3\times10^{-5}$ | 19 |
| $WSe_2$ | 450 | 0 | - | 0.1 | - | $2.2\times10^{13}$ | $5.5\times10^{-7}$ | 20 |
| $MoTe_2$ | 635 | 0 | - | 0.5 | - | $1.6\times10^{12}$ | 0.003 | 21 |
| BP | 1450 | 0 | - | 1.06 | - | $1.3\times10^{11}$ | $3.6\times10^{-4}$ | 22 |

[a)] Red fill indicates phototransistors with uniform channel, and blue fill indicates photodiodes without bias. The performance metrics selected from the references are based on the principle of maximum responsivity.
[b)] Response time is defined as the maximum of rise time and fall time, both extracted between 10% and 90% of the maximum photocurrent.

**Table S2.** Parameters for gain calculation of the main PVPT device under 532 nm illumination

| Parameter | Value | Calculation: Performance | Formula | Result |
|---|---|---|---|---|
| $I_{ph}$ | 1.25 μA | | | |
| $A$ | 62.8 μm² | $R$ | $\frac{I_{ph}}{P_{in}\cdot A}$ | $1.14\times10^5$ A/W |
| $P_{in}$ | $0.016\ mW/cm^2$ | $EQE$ | $\frac{Rhc}{e\lambda}$ | $2.65\times10^7$% |
| $\eta$ | 1.7‰ | $G$ | $\frac{I_{ph}/e}{\Phi_{in}\eta} = \frac{EQE}{\eta}$ | $1.55\times10^8$ |

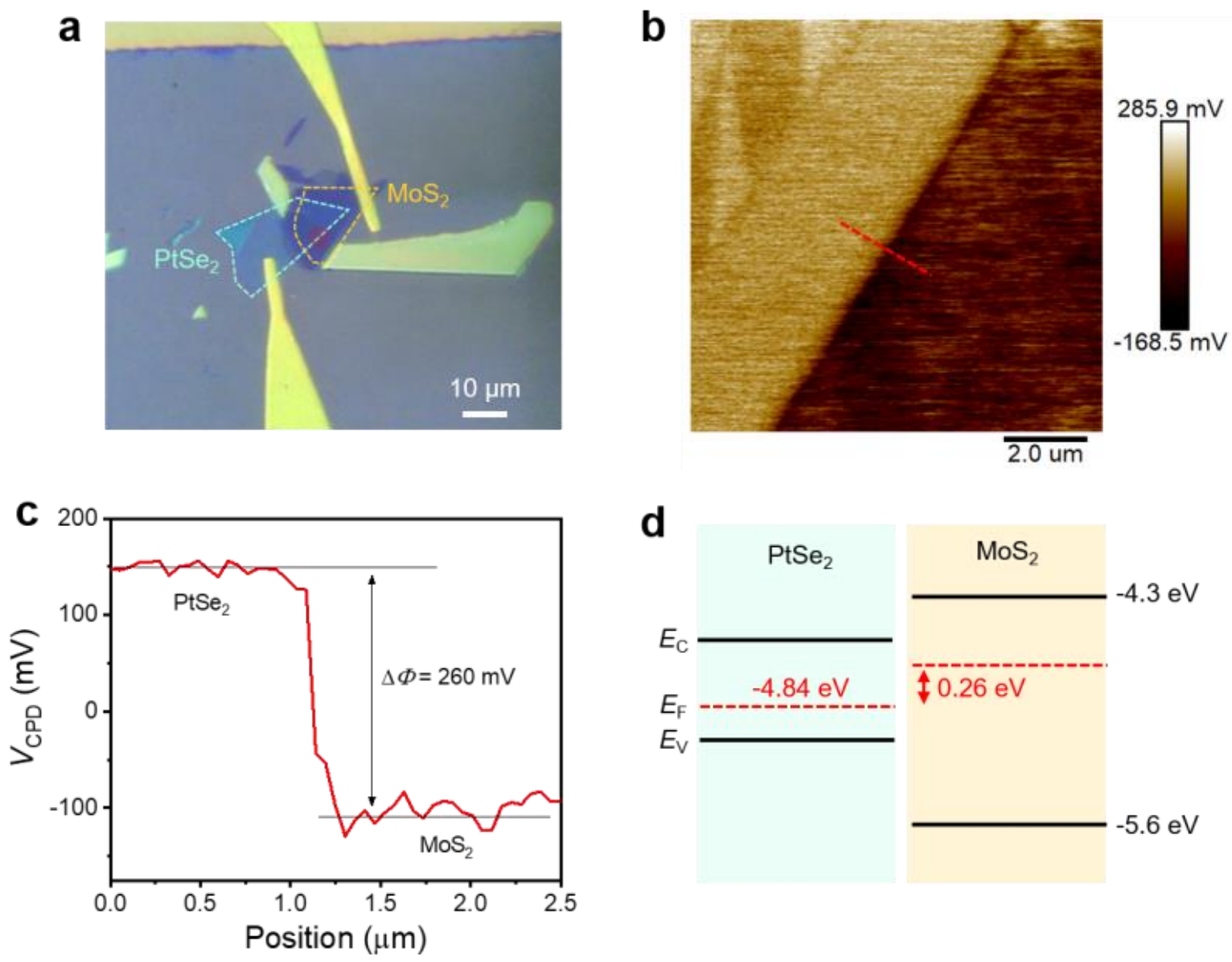

**Figure S1.** (a) Optical image of a $PtSe_2/MoS_2$ heterostructure on an hBN sheet. (b) KPFM image of the red box in the optical image of (a). (c) Contact potential difference ($V_{CPD}$) profile along the red dashed line in (b). (d) the band diagram of $PtSe_2$ and $MoS_2$ before contact.

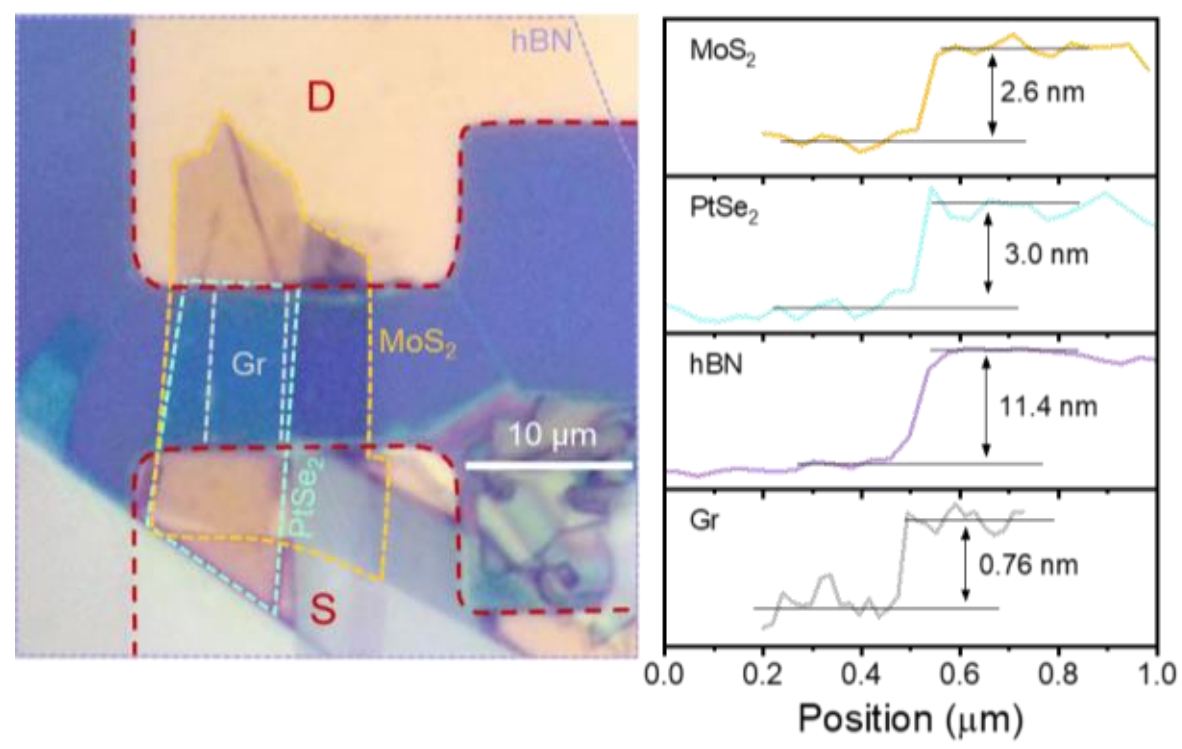

**Figure S2.** Optical image (left) and AFM profile (right) of PVPT device used for photocurrent scanning.

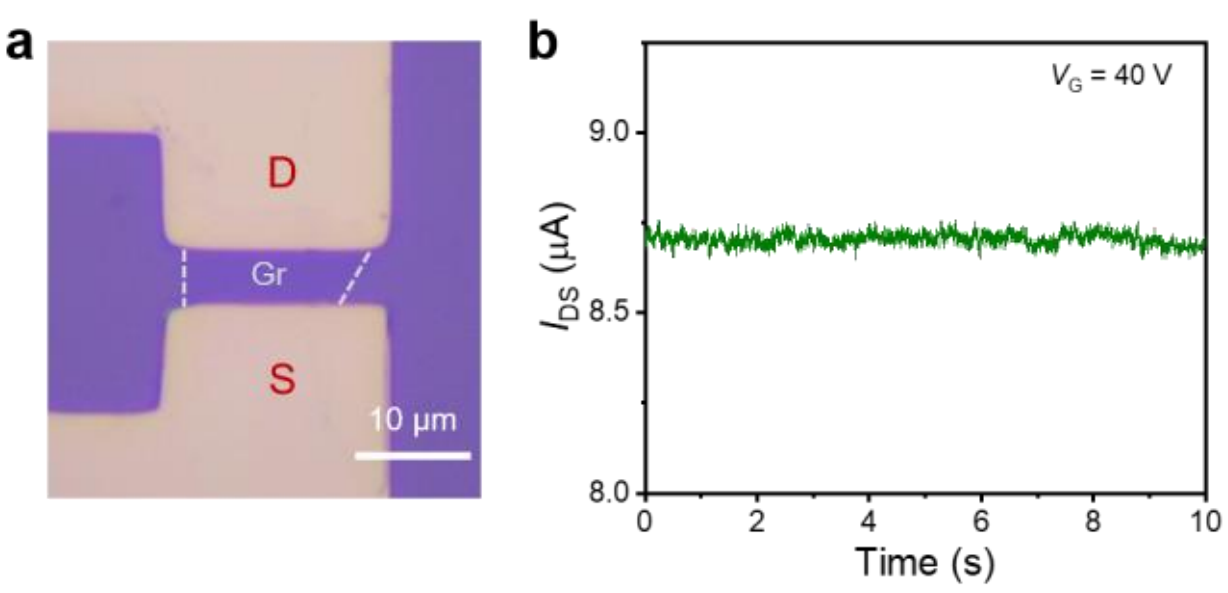

**Figure S3**. Photoresponse of a graphene phototransistor under 532 nm laser illumination. (a) Optical image of the device. (b) Transient photocurrent under modulated illumination (1 Hz) with $P_{\mathrm{in}} = 8.7$ mW/cm$^2$ at $V_{\mathrm{DS}} = 5$ mV and $V_{\mathrm{G}} = 40$ V.

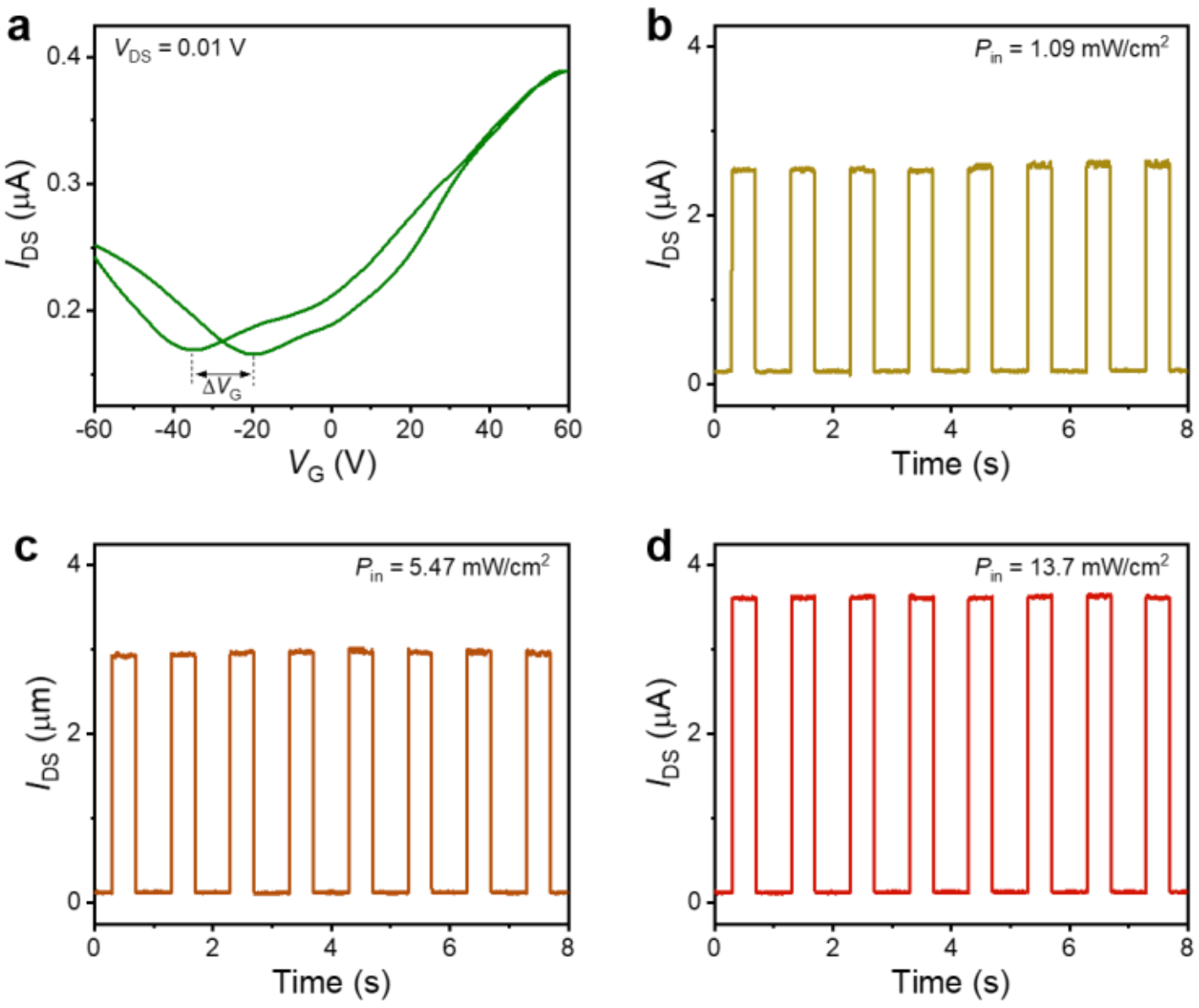

**Figure S4**. Optoelectronic performance of a PVPT device (see Figure S5) at $V_{DS} = 10$ mV. (a) Dual-sweeping transfer curves, where $\Delta V_G = 15.6$ V. (b-d) Transient photocurrent measured under modulated illumination (532 nm, 1 Hz) at $V_G = 40$ V for laser power densities of $P_{in} = 1.09$, 5.47, and 13.7 mW/cm$^2$, respectively.

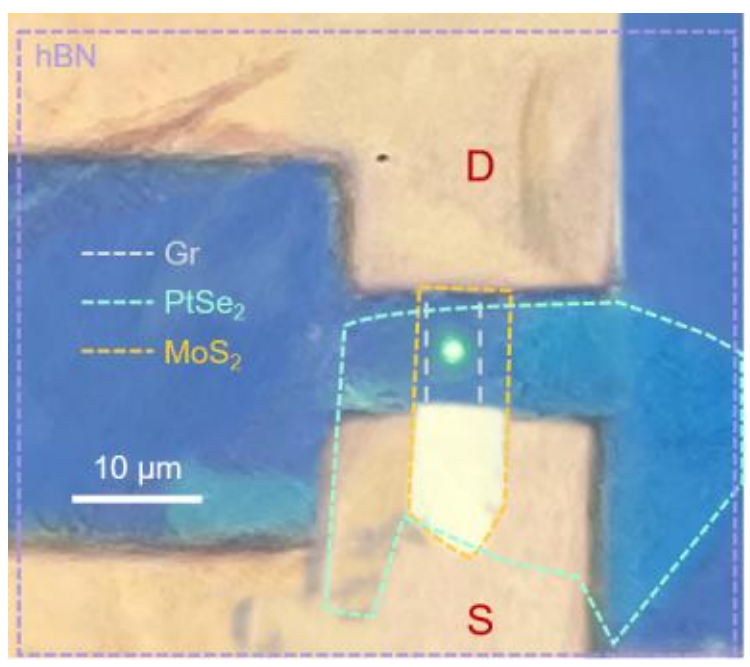

**Figure S5.** Optical image of a PVPT device used for dual-sweeping transfer characterization and temperature-dependent photodetection. The focused 532 nm laser spot (diameter of ~ 3 μm) is shown for the size comparison with the Gr channel.

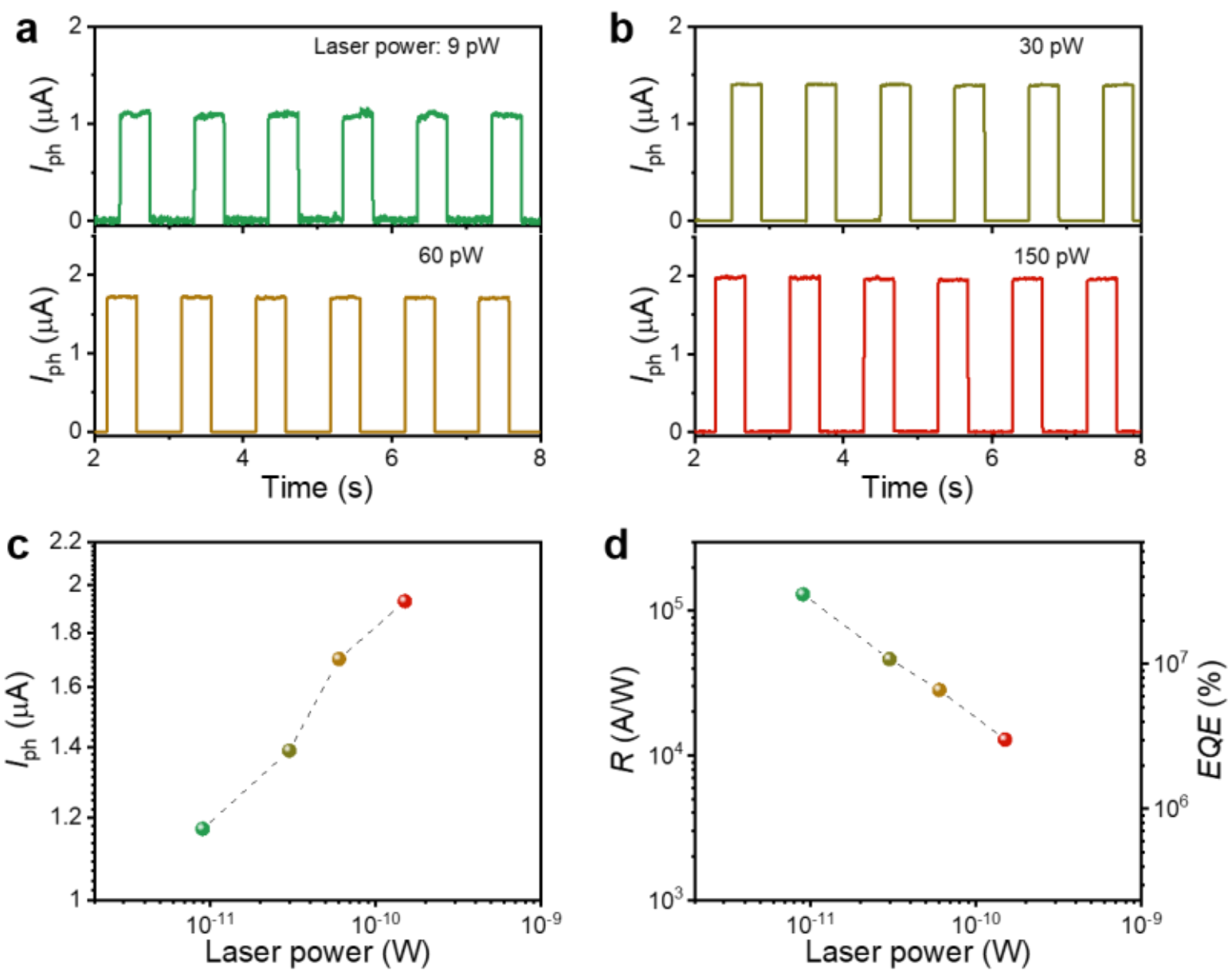

**Figure S6.** Photodetection performance of a PVPT device (Figure 1d) illuminated by a focused 532 nm laser spot (diameter of ~ 3 μm). (a-b) Transient photocurrent at $V_G = 40$ V and $V_{DS} = 1$ mV for the laser power of 9, 30, 60, and 150 pW, respectively. (c) Photocurrent versus laser power. (d) Responsivity ($R_{max} = 1.3 \times 10^5$ A/W) and external quantum efficiency ($EQE_{max} = 3.04 \times 10^7$) as a function of laser power.

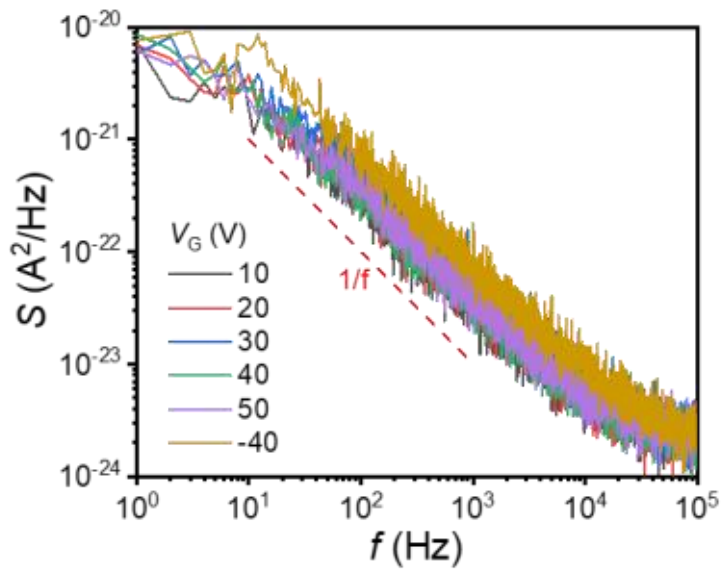

**Figure S7.** Spectral noise density of the main device (Figure 1d) at different gate voltages, measured under dark conditions.

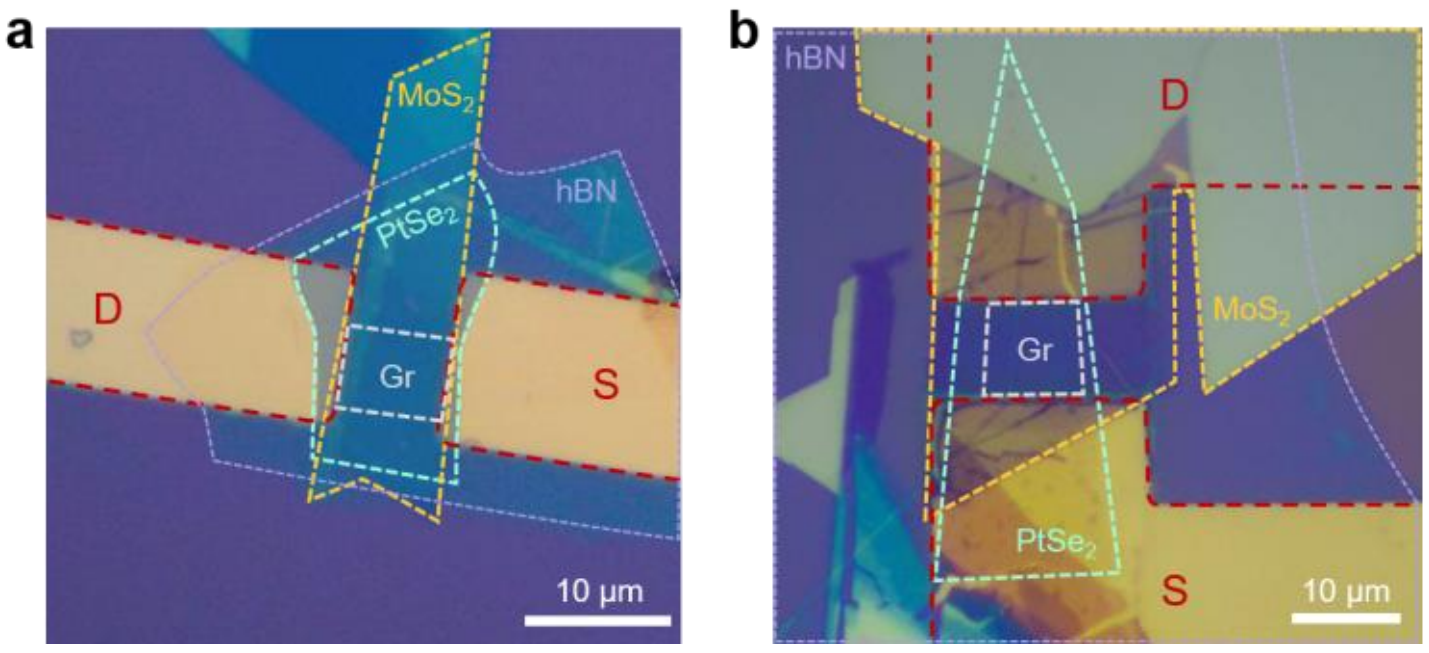

**Figure S8.** Optical images of the PVPT devices used for high-speed testing when the laser is modulated by (a) TTL and (b) AOM, respectively.

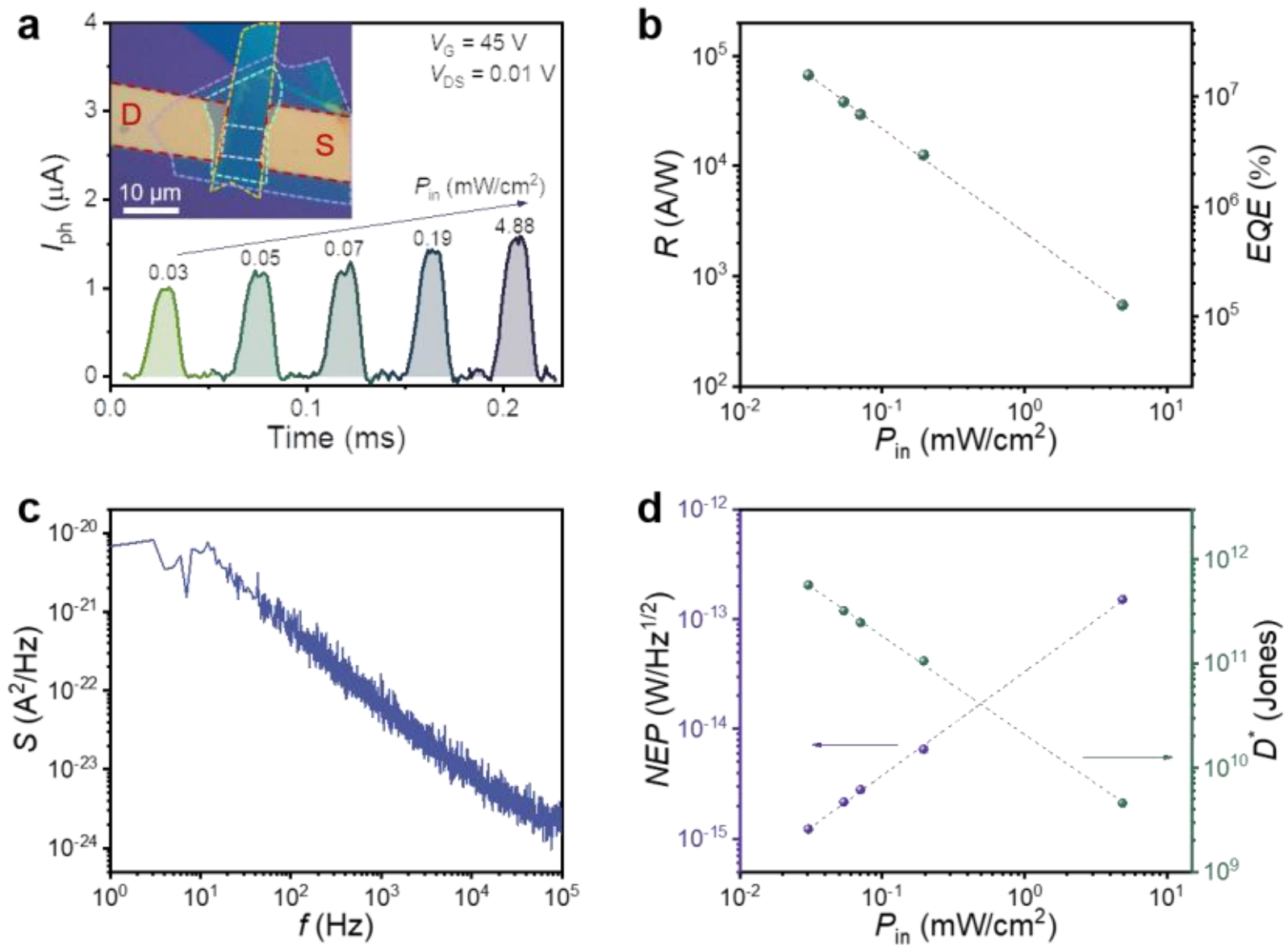

**Figure S9.** Photodetection performance of the PVPT device under 532 nm illumination with TTL modulation at 25 kHz. (a) Transient photocurrent measured at varying incident power density $P_{\text{in}}$. Inset: optical image of the device. (b) $R$ and $EQE$ as functions of $P_{\text{in}}$. (c) Spectral noise density measured under dark conditions at the same voltages. (d) $NEP$ and $D^*$ as functions of $P_{\text{in}}$. All measurements were performed at $V_{\text{G}} = 45$ V and $V_{\text{DS}} = 0.01$ V.

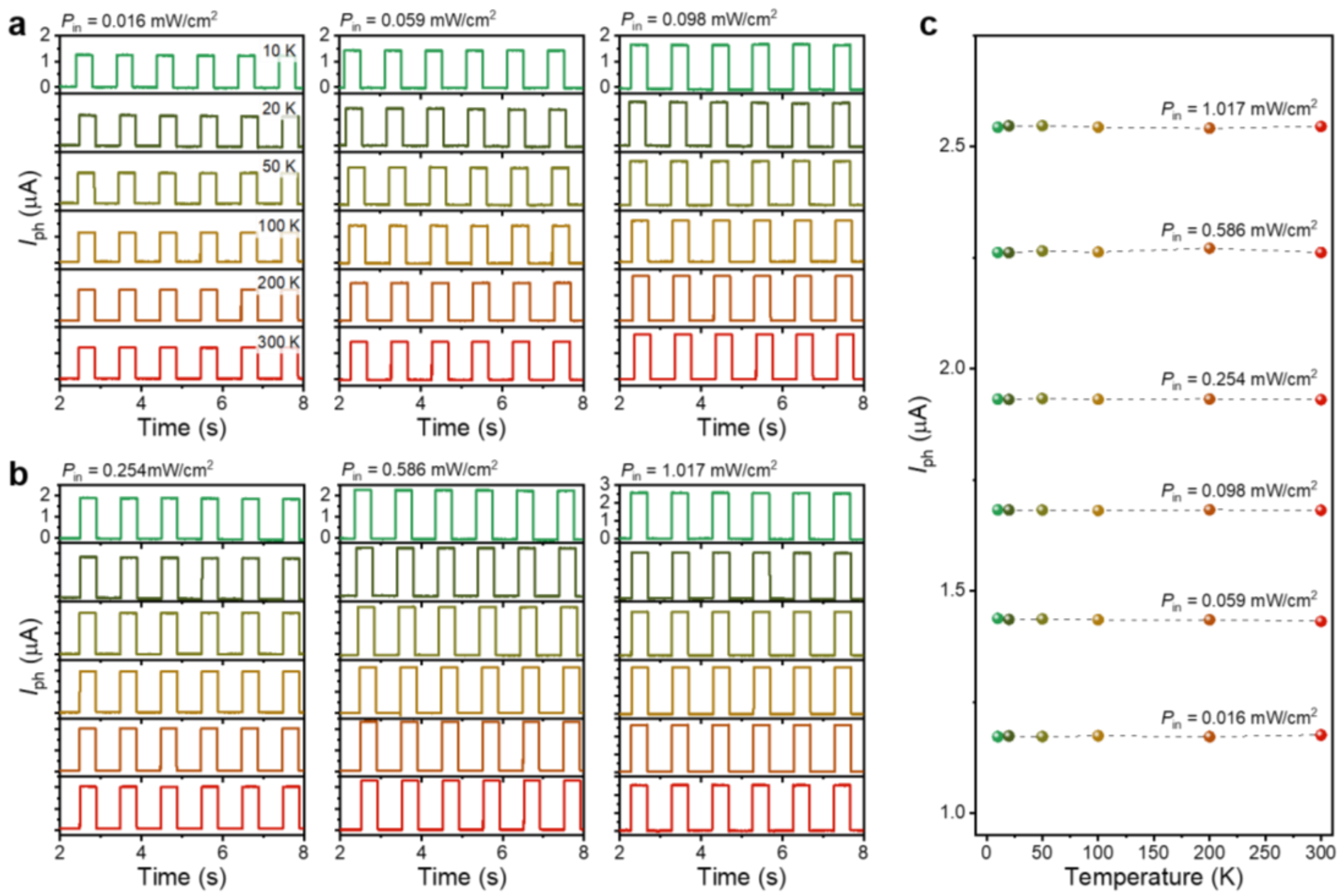

**Figure S10.** Temperature-dependent photoresponse of the PVPT device (see Figure S5) under 532 nm modulated illumination (1 Hz) at $V_{\text{G}} = 30$ V and $V_{\text{DS}} = 10$ mV. (a-b) Transient photocurrent at various temperatures for different $P_{\text{in}}$. (c) Extracted photocurrent as a function of temperature under different $P_{\text{in}}$.

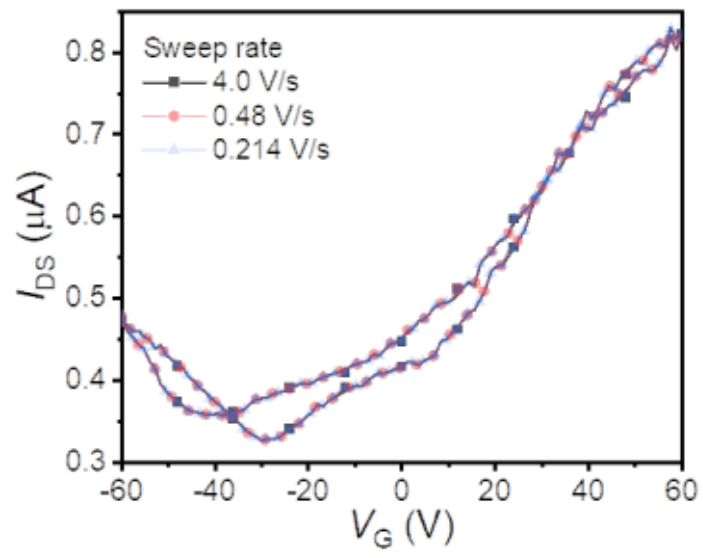

**Figure S11** Dual-sweeping transfer curves of a PVPT device (Figure S5) at different sweep rates at $V_{\mathrm{DS}} = 10$ mV.

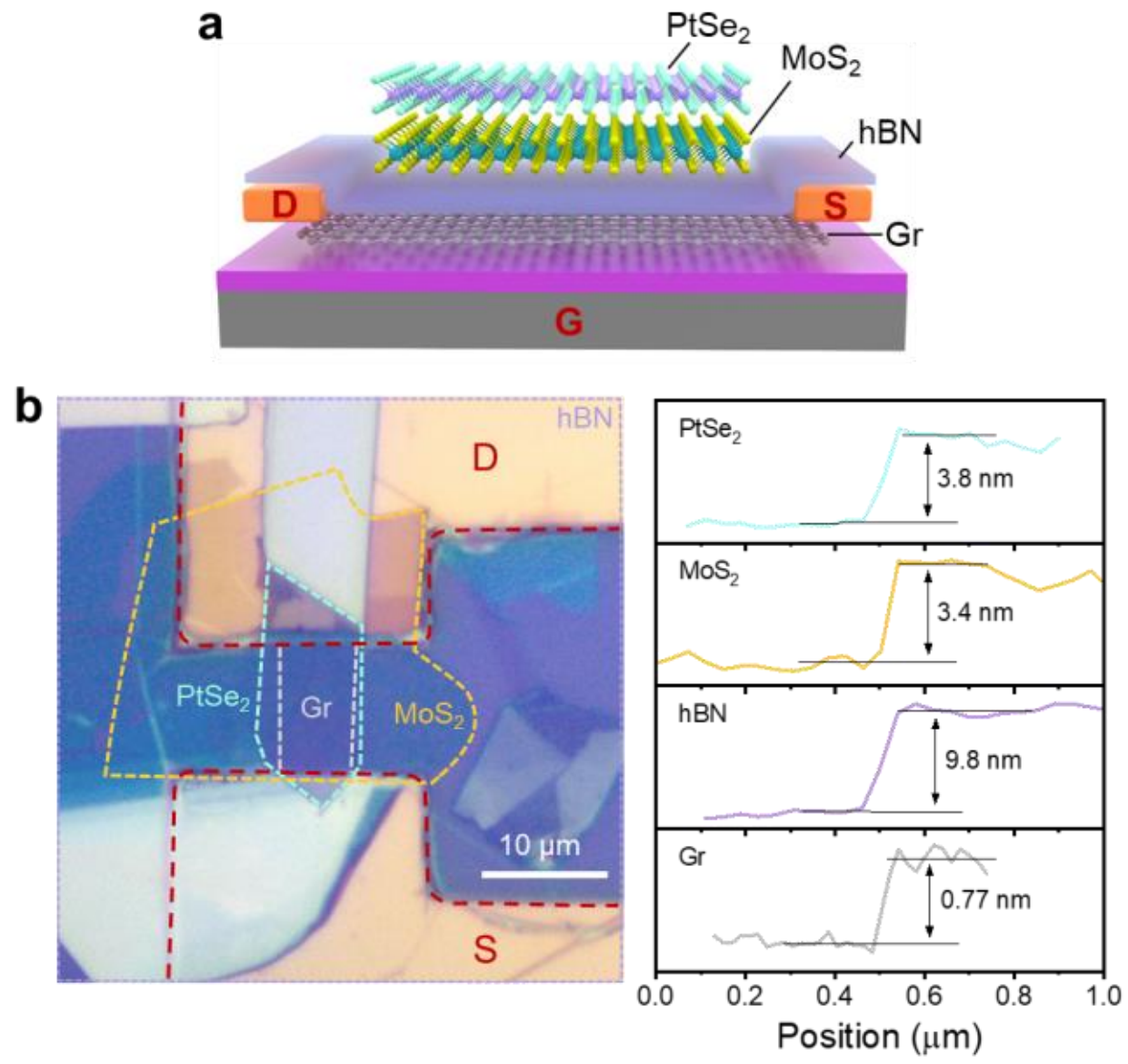

**Figure S12. Structure of the PVPT device gated by $PtSe_2/MoS_2$ heterostructure.** (a) Schematic diagram, (b) optical image (left) and AFM profile (right) of the device.

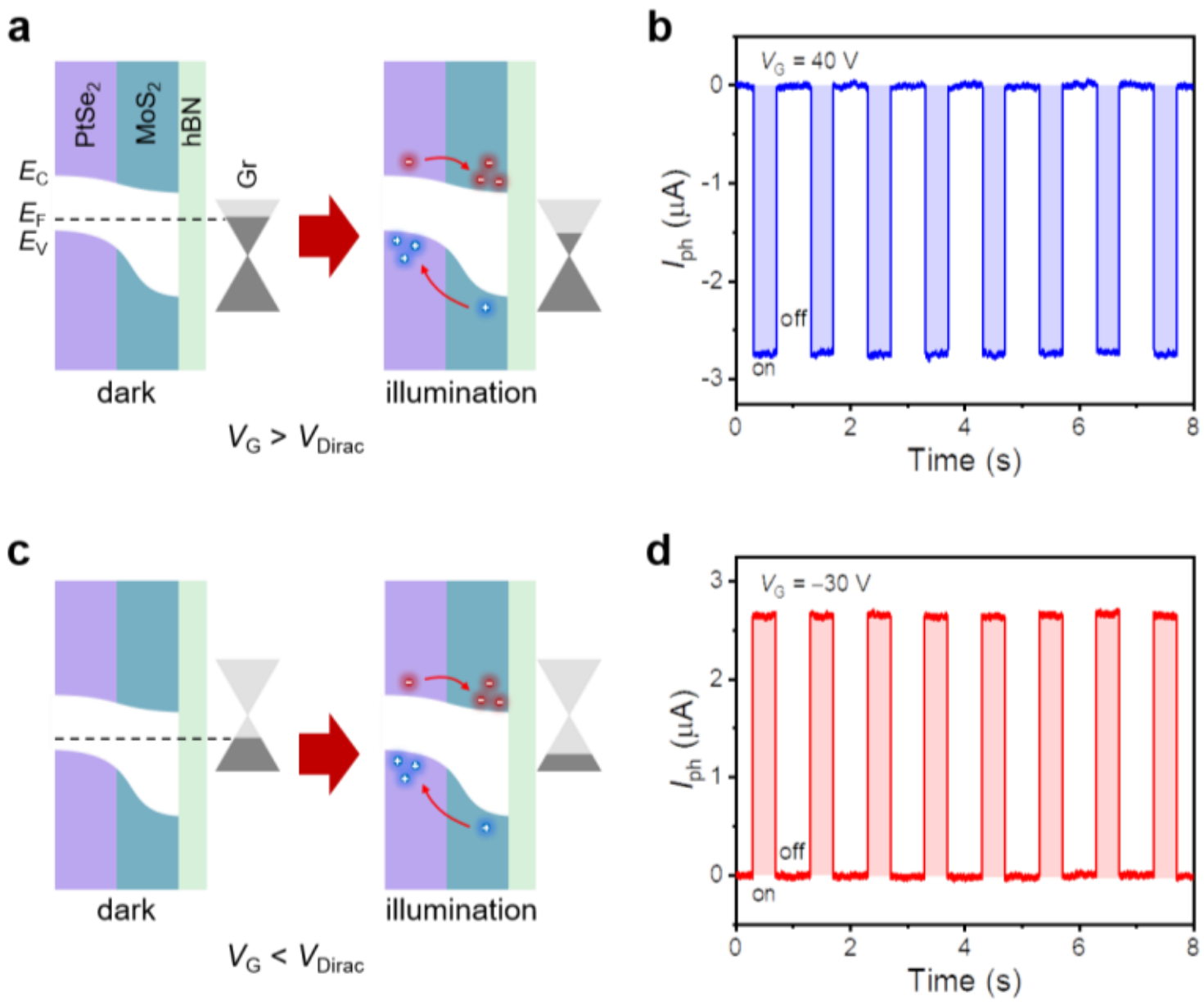

**Figure S13. Photodetection mechanism of the device gated by $PtSe_2$/$MoS_2$ heterostructure.** (a) Band structure diagrams of the $PtSe_2$/$MoS_2$-gated Gr phototransistor under dark and illumination conditions for electron transport at $V_G > V_{Dirac}$. (b) Negative photoresponse when $V_G = 40$ V. (c) Band structure diagrams of the device under dark and illumination conditions for the hole transport when $V_G < V_{Dirac}$. (d) Positive photoresponse when $V_G = -30$ V. The excitation is a 532 nm laser with $P_{in} = 9.78$ mW/cm$^2$.

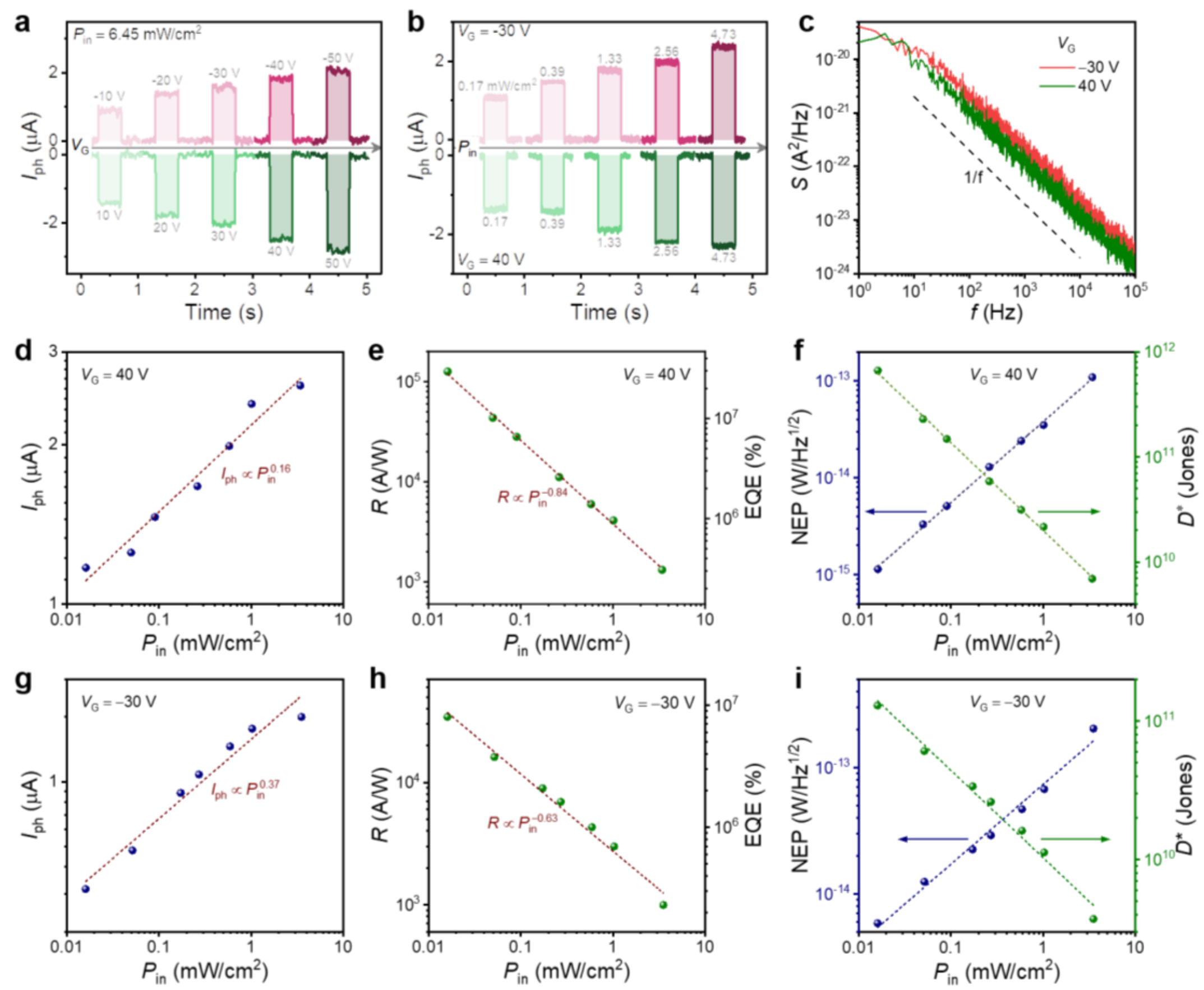

**Figure S14. Photodetection performances of the device gated by $PtSe_2/MoS_2$ heterostructure.** (a) Transient photocurrent for different $V_G$ at $P_{in} = 6.45$ mW/cm$^2$. (b) Transient photocurrent for different $P_{in}$ at $V_G = -30$ and 40 V. (c) Spectral noise density of the main device mearsured under dark conditions at different gate voltages. (d-f) $I_{ph}$, $R$, $EQE$, $NEP$, and $D^*$ with respect to $P_{in}$ when $V_G = 40$ V. (g-i) corresponding performance with respect to $P_{in}$ when $V_G = -30$ V.

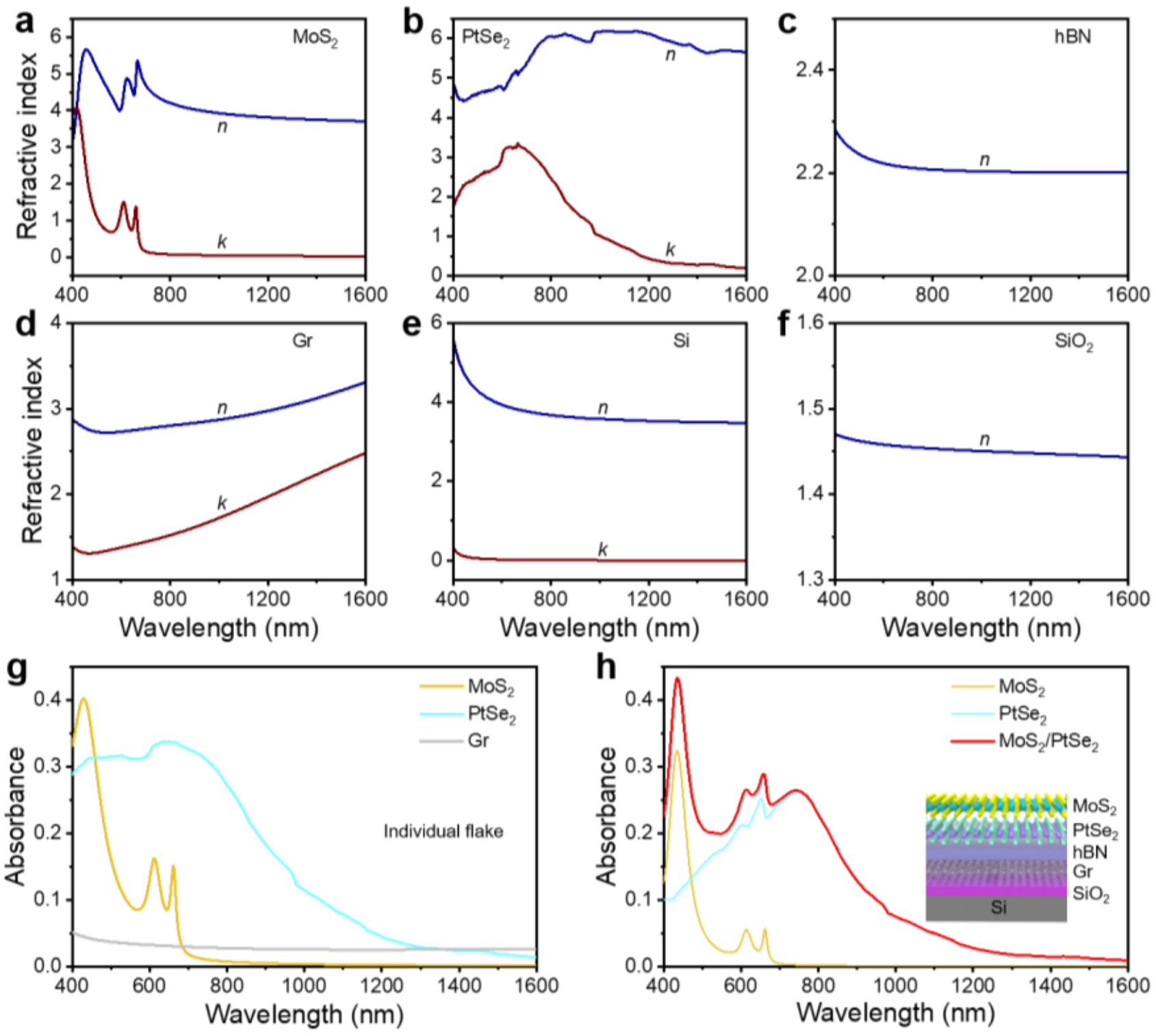

**Figure S15.** (a-f) Refractive indices of different flakes as a function of wavelength, obtained from the Refractive Index Database (https://refractiveindex.info/). (g) Normal-incidence absorption spectra of individual single flakes. (h) Absorption spectra of the complete flake stack in the PVPT device, with the Si substrate treated as semi-infinite. All absorption spectra were calculated using transfer matrix method, with layer thicknesses taken from the main device (Figure 1d): $MoS_2$ (1.43 nm), $PtSe_2$ (1.73 nm), hBN (10.3 nm), Gr (0.43 nm), and $SiO_2$ (285 nm).

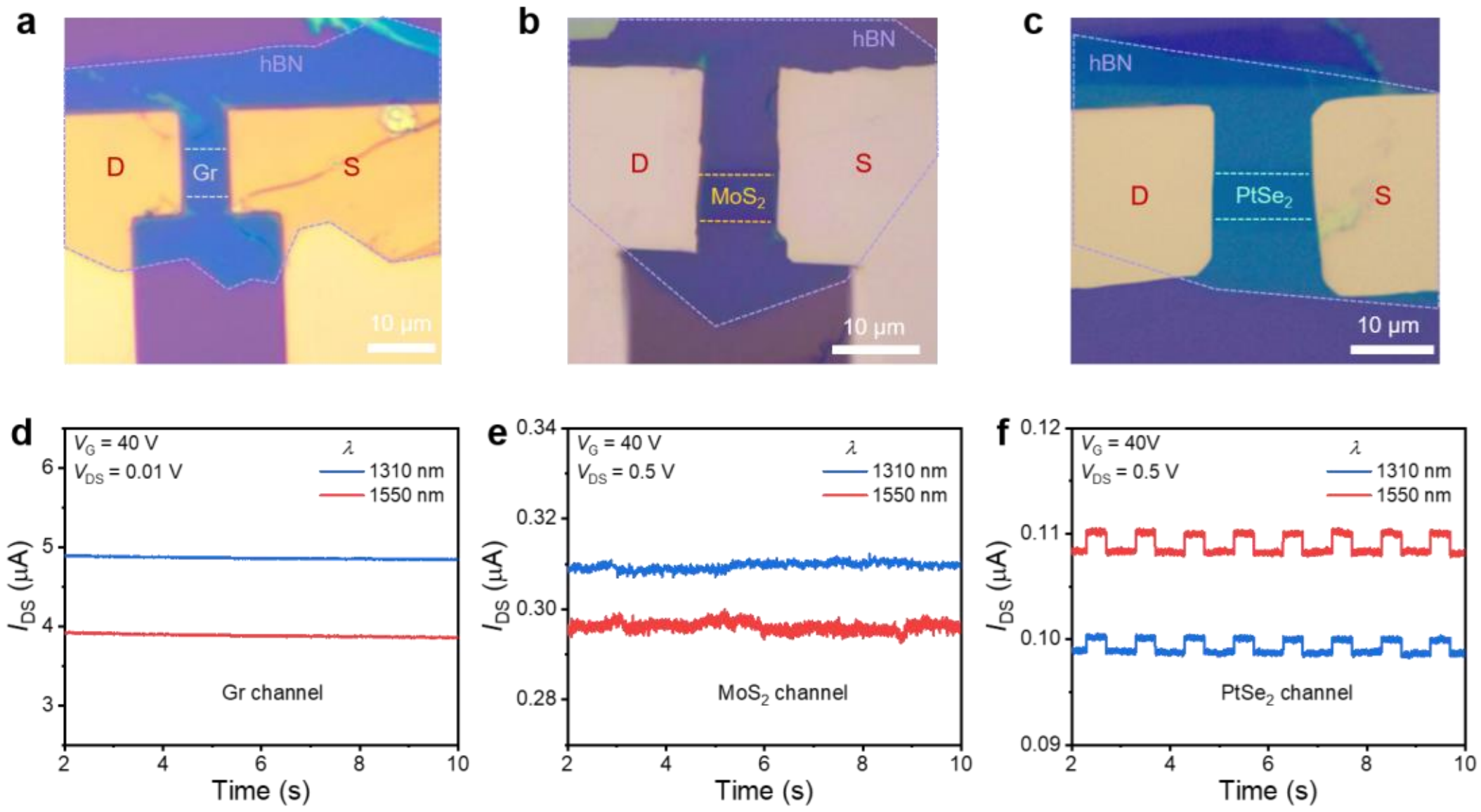

**Figure S16** Near-infrared photoresponse of transistors with different channel materials under 1 Hz modulation. Optical images of devices with (a) Gr, (b) $MoS_2$, and (c) $PtSe_2$ channels, respectively. (d-f) Transient photocurrent responses of the corresponding devices under 1310 nm and 1550 nm illumination. The incident power densities were 6.67 $mW/cm^2$ for 1310 nm and 7.59 $mW/cm^2$ for 1550 nm.

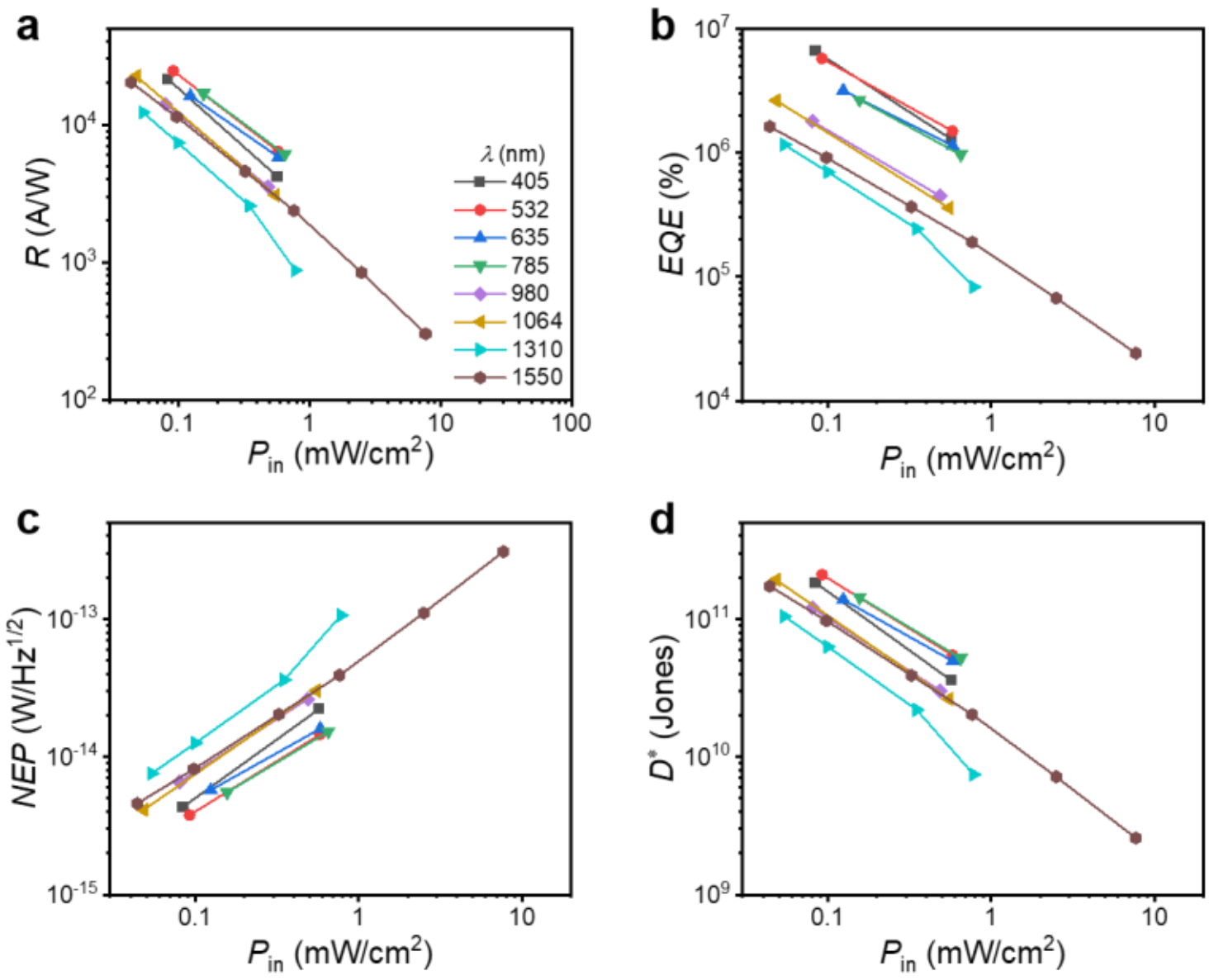

**Figure S17. Photodetection performance of the main device for different wavelengths.** (a) Responsivity, (b) external quantum efficiency, (c) noise equivalent power, and (d) detectivity with respect to incident power density.

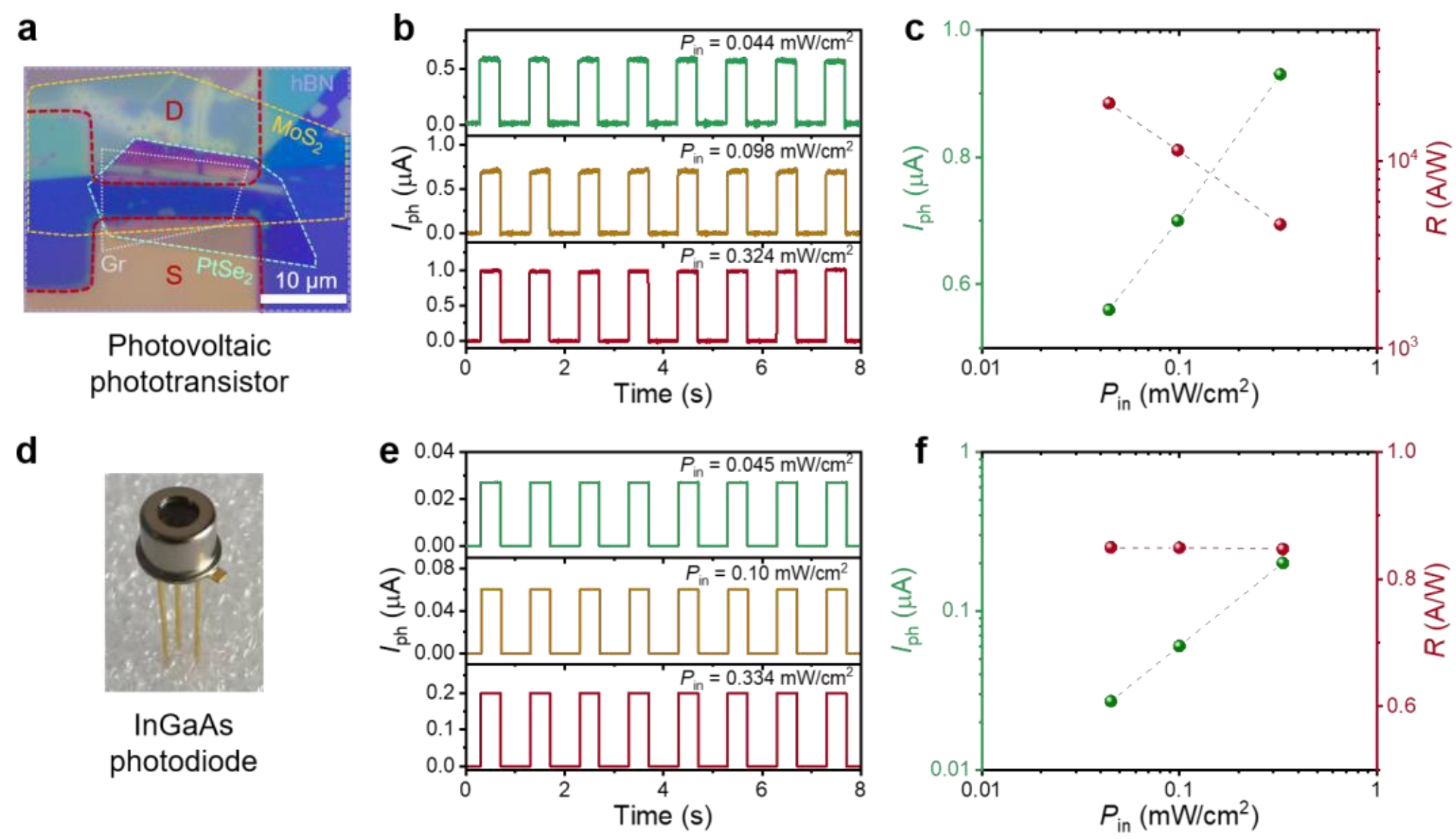

**Figure S18.** (a-c) Optical image, transient photocurrent, and responsivity of a PVPT under 1550 nm illumination at $V_G = 40$ V and $V_{DS} = 1$ mV. (d-f) Optical image, transient photocurrent, and responsivity of a commercial InGaAs photodiode (diameter 300 μm) under 1550 nm illumination at the bias voltage of 1 mV.

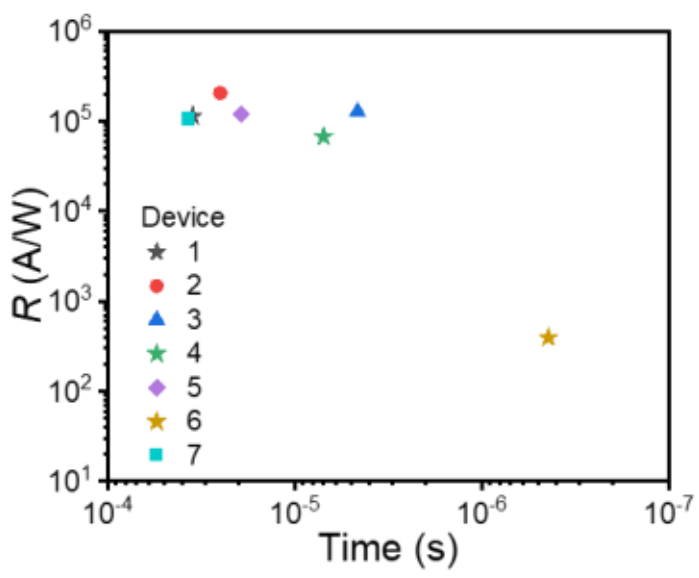

**Figure S19.** Performance summary of six representative PVPT devices under 532 nm illumination. The figure references corresponding to each device are as follows: Device 1 (Figures 1d, 3, 4b, and S17); Device 2 (Figures 2i, 2l, and S2); Device 3 (Figures S4, S5, S6, S10, and S11); Device 4 (Figures 4c, S8a, and S9); Device 5 (retest of Device 1) (Figures S18a-S18c); Device 6 (Figures 4d and S8b); Device 7 (Figures S12-S14). It should be noted that the values presented here are not necessarily optimal; the intrinsic response speed is expected to be faster, and higher responsivity can be achieved under weaker illumination.